\documentclass[prb,showpacs,twocolumn,amsmath,amssymb,superscriptaddress]{revtex4-1}
\usepackage[colorlinks=true, citecolor=blue, urlcolor=blue ]{hyperref}
\usepackage{graphicx}
\usepackage{subfigure}
\usepackage{grffile}
\usepackage{bm}
\usepackage[version=3]{mhchem}
\usepackage{color}
\usepackage{hyperref}
\usepackage[normalem]{ulem}

\newcommand{\ket}[1]{{\vert #1\rangle}}

\newcommand{\braket}[2]{\langle#1\vert#2\rangle}
\newcommand{\eval}[3]{\langle#1\vert#2\vert#3\rangle}

\newcommand{\1}{\mbox{\bf 1}}

\begin{document}

\title{Time-resolved single-particle spectrum of the one-dimensional extended Hubbard model after interaction quenches}

\author{Yong-Guang Su}
\affiliation{Institute of Ultrafast Optical Physics, Department of Applied Physics $\&$ MIIT Key Laboratory of Semiconductor Microstructure and Quantum Sensing, Nanjing University of Science and Technology, Nanjing 210094, China}

\author{Ruifeng Lu}
\affiliation{Institute of Ultrafast Optical Physics, Department of Applied Physics $\&$ MIIT Key Laboratory of Semiconductor Microstructure and Quantum Sensing, Nanjing University of Science and Technology, Nanjing 210094, China}

\author{Hantao Lu}
\affiliation{School of Physical Science and Technology $\&$ Lanzhou Center for Theoretical Physics, Key Laboratory of Theoretical Physics of Gansu Province, Lanzhou University, Lanzhou 730000, China}

\author{Can Shao}
\email{shaocan@njust.edu.cn}
\affiliation{Institute of Ultrafast Optical Physics, Department of Applied Physics $\&$ MIIT Key Laboratory of Semiconductor Microstructure and Quantum Sensing, Nanjing University of Science and Technology, Nanjing 210094, China}

\date{\today}

\begin{abstract}
We investigate the non-equilibrium dynamics of the one-dimensional extended Hubbard model after interaction quenches. In strong-coupling regime with large on-site interaction, the ground states of this model with small and large nearest-neighbor interactions are in spin-density-wave and charge-density-wave phases, respectively. Combining twisted boundary conditions with the time-dependent Lanczos method, we obtain snapshots of the time-dependent single-particle spectrum after quenches. We find that for quench within the same phase, the single-particle spectrum becomes close to that of the quenched Hamiltonian immediately after the quench. While for quench across the critical point, the afterward evolution process depends mainly on the distribution of the initial state among the eigenstates of the quenched Hamiltonian. Our finding may serve as a way to detect the phase transition in ultracold atom systems with interactions.

\end{abstract}


\maketitle

\section{Introduction}
\label{sec1}
Non-equilibrium processes of the interacting quantum many-particle systems have attracted much attention and been widely studied in the past few years~\cite{Polkovnikov11,Jacek10,Cazalilla_2010}.
One example is the interaction quench in isolated systems, where the initial state is the ground state before quench and its evolution is then governed by the quenched Hamiltonian.
It is experimentally accessible in ultracold atoms, which are trapped on optical lattices and almost isolated from the environment~\cite{Kinoshita2006, Gring12, Trotzky2012}.
A generic closed quantum system is expected to thermalize after non-equilibrium dynamics, with local observables being accurately described in the end by the equilibrium statistical mechanics~\cite{Rigol2008}. From the viewpoint of the eigenstate thermalization hypothesis (ETH)\cite{Luca16, Deutsch_2018}, individual energy eigenstates behave like a statistical ensemble in the sense that a local subsystem can reach a stationary state and thermalize\cite{Calabrese2020}.
However, some specific quantum systems have been found to evolve into nonthermal states with much more initial information preserved than the usual thermal ones~\cite{Rigol07, Rigol06, Cazalilla06, Iucci09, Kollar08, Eckstein08, Fioretto_2010}.
In addition, a thermalization process may rely on details, e.g., the strength of the quench~\cite{Roux09} and the distance of the system away from the integrable point~\cite{Kollath07, Moeckel08, Moeckel09, Rigol09}.
Many questions remain open at present, e.g., what are the characteristics of the steady state after quench and when will it relax to a thermal state?
The question has been extensively discussed and is believed to be relevant to the quantum ergodicity and its breaking~\cite{Polkovnikov11}.

In recent years, the time-resolved optical spectroscopy, such as the time- and angular-resolved photoemission spectroscopy (trARPES), has been widely applied to study the non-equilibrium dynamics of materials~\cite{Rohwer2011,Petersen11,Perfetti06,Perfetti_2008,Avigo_2016,Wu_2021,Schmitt_2011}.
It has the potential to unravel the complicated couplings between different degrees of freedom on different time scales to some extent.
Experimental data on trARPES are often compared with the time-resolved single-particle spectral functions in theory, and the latter can be obtained by combining the non-equilibrium Green's functions, i.e. Keldysh formalism, with the dynamical mean-field theory~\cite{Aoki14, Nosarzewski17, Kemper15, Kemper17}. An alternative way to describe the dynamical response of correlated systems out of equilibrium relies on the time-dependent wave functions, which can be secured using various numerical methods, e.g., the exact diagonalization\cite{ShaoCan2022, ShaoCan2019, ShaoCan2016}. Due to the finite size used in the numerical simulations, the time-dependent single-particle spectral function can be merely resolved at certain points of momentum~\cite{Kanamori09}. While the application of twisted boundary conditions makes it posibble to increase the momentum resolution of the spectrum~\cite{Tsutsui96, Tohyama04}, and a time-dependent version has also been proposed in Ref.~\onlinecite{Shao20}.

In this paper, inspired by a recent proposal of using a quantum gas microscope to experimentally access the momentum- and energy-resolved spectral function~\cite{Bohrdt18}, we utilize the time-dependent single-particle spectral function to investigate the quench dynamics of the one-dimensional extended Hubbard model (1D EHM) at half filling. Such quench process can also be realized in cold atom systems, where the time-resolved quantities, including the spectral functions, are ready to be measured.
In the 1D EHM, both the on-site and nearest-neighbor repulsions (denoted as $U$ and $V$) are taken into account, and this electronic model exhibits rich phases\cite{Emery}.
However, in the strong-coupling regime (with large on-site interactions), the ground-state phase diagram of the model is relatively simple: it is divided into the spin-density wave (SDW) and charge-density wave (CDW) phases\cite{Tsuchiizu02, Ejima07}. We find that when the interaction quench happens within the same phase, the time-dependent single-particle spectrum changes to the counterpart of the quenched Hamiltonian immediately after the quench. On the other hand, if the quench crosses different phases, the evolution of the spectrum depends mainly on the overlaps between the initial state and the eigenstates of the quenched Hamiltonian. Our finding may be used to detect the phase transition in ultracold atom systems.

The rest of the paper is organized as follows. In Sec.~\ref{sec_model}, we introduce the model, method and relevant properties. By analyzing the time-dependent single-particle spectrum, we study the non-equilibrium dynamics of quantum quenches in Sec.~\ref{sec_pump}. The conclusion is given in Sec.~\ref{sec_conclusion}.

\section{Model and Method}\label{sec_model}

The Hamiltonian of 1D EHM reads
\begin{eqnarray}
H&=&-t_h\sum_{i,\sigma}\left(c^{\dagger}_{i,\sigma} c_{i+1,\sigma}+\text{H.c.}\right)+U\sum_{i}\left(n_{i,\uparrow}-\frac{1}{2}\right)\nonumber\\
&&\times\left(n_{i,\downarrow}-\frac{1}{2}\right)
 +V\sum_{i}\left(n_{i}-1\right)\left(n_{i+1}-1\right),
\label{H}
\end{eqnarray}
where $c^{\dagger}_{i,\sigma}$ ($c_{i,\sigma}$) creates (annihilates) an electron at site $i$ with spin $\sigma=\uparrow,\downarrow$, and $n_i= n_{i,\uparrow}+ n_{i,\downarrow}$ is the number operator of electrons. $t_h$ is the hopping constant; $U$ and $V$ are the on-site and nearest-neighbor Coulomb repulsion strengths, respectively.
We use units with $a_0=e=\hbar=c=1$ in the rest of the paper, where $a_0$, $e$, $\hbar$ and $c$ are the lattice constant, the elementary charge, the reduced Planck constant and the speed of light, respectively. In these units, ${t_h}^{-1}$ and $t_h$  are set to be the unit of time and energy, respectively.

In equilibrium, ARPES is often compared with the single-particle spectral function $I(k,\omega)$ .
If the standard periodic boundary condition is used for a 1D $L$-site lattice, the allowed momenta in the first Brillouin zone satisfy $k_0=2\pi l/L$, where $l=0,1,...,L-1$. In order for $k$ to cover the full Brillouin zone, the twisted boundary condition is adopted\cite{Poilblanc91, Tsutsui96, Tohyama04}. For a momentum $k=k_{0}+ \kappa$, where $\kappa$ is arbitrary, imposing the twist is equivalent to the following transformation:
\begin{eqnarray}
c^{\dagger}_{i,\sigma} c_{i+1,\sigma} \rightarrow
e^{i{\kappa}}c^{\dagger}_{i,\sigma} c_{i+1,\sigma}.
\end{eqnarray}
The single-particle spectral function $I(k,\omega)$ at zero temperature can be written as\cite{Tohyama04}:
\begin{eqnarray}
I(k,\omega)=I_{+}(k,\omega)+I_{-}(k,\omega),
\label{A}
\end{eqnarray}
with
\begin{eqnarray}
I_{+}(k,\omega)=\sum_{m,\sigma}|\langle\Psi_m^{\kappa}|c_{k_0,\sigma}^{\dag}|\Psi_0^{\kappa}\rangle|^{2} \delta(\omega+(E_m^{\kappa}-E_0^{\kappa})) \nonumber \\
=-\frac{1}{\pi}\text{Im}\left(\langle\Psi_0^{\kappa}|c_{k_0,\sigma}\frac{1}{\omega+(H^{\kappa}-E^{\kappa}_0)}c_{k_0,\sigma}^{\dag}|\Psi_0^{\kappa}\rangle\right),
\label{A+}
\end{eqnarray}
\begin{eqnarray}
I_{-}(k,\omega)=\sum_{m,\sigma}|\langle\Psi_m^{\kappa}|c_{k_0,\sigma}|\Psi_0^{\kappa}\rangle|^{2} \delta(\omega-(E_m^{\kappa}-E_0^{\kappa})) \nonumber \\
=-\frac{1}{\pi}\text{Im}\left(\langle\Psi_0^{\kappa}|c_{k_0,\sigma}^{\dag}\frac{1}{\omega-(H^{\kappa}-E^{\kappa}_0)}c_{k_0,\sigma}|\Psi_0^{\kappa}\rangle\right).
\label{A-}
\end{eqnarray}
Here $I_{+}$ ($I_{-}$) is the electron-addition (electron-removal) spectral function. The $\ket{\Psi_0^{\kappa}}$ and $\ket{\Psi_m^{\kappa}}$ represent the ground state with the energy $E^{\kappa}_0$ and the final state with $E^{\kappa}_m$, respectively, for a given $\kappa$.

As detailed in Ref.~\onlinecite{Shao20}, the time-dependent version of the single-particle spectral function can be obtained through the following process. The Hamiltonian changes after the interaction quench and $\ket{\Psi^{\kappa}_0}$ is usually no longer an eigenstate of the quenched Hamiltonian $H^{\kappa}(t)$. The time-evolving wave function, denoted as $\ket{\Psi^{\kappa}(t)}$, can be calculated by using the standard time-dependent Lanczos method~\cite{Prelovsek}.
Then, by the substitution of $\ket{\Psi_0^{\kappa}}$ in Eqs.~(\ref{A+}) and (\ref{A-}) with $\ket{\Psi^{\kappa}(t)}$, and $H^{\kappa}$ with $H^{\kappa}(t)$, the time-dependent spectral function $I_{\pm}(k,\omega,t)$, can be obtained. Take notice that $E_0^{\kappa}$ in Eqs.~(\ref{A+}) and (\ref{A-}) should also be replaced by $E^{\kappa}(t)=\eval{\Psi^{\kappa}(t)}{H^{\kappa}(t)}{\Psi^{\kappa}(t)}$ in the time-dependent calculations.

\begin{figure}[t]
\centering
\includegraphics[width=0.5\textwidth]{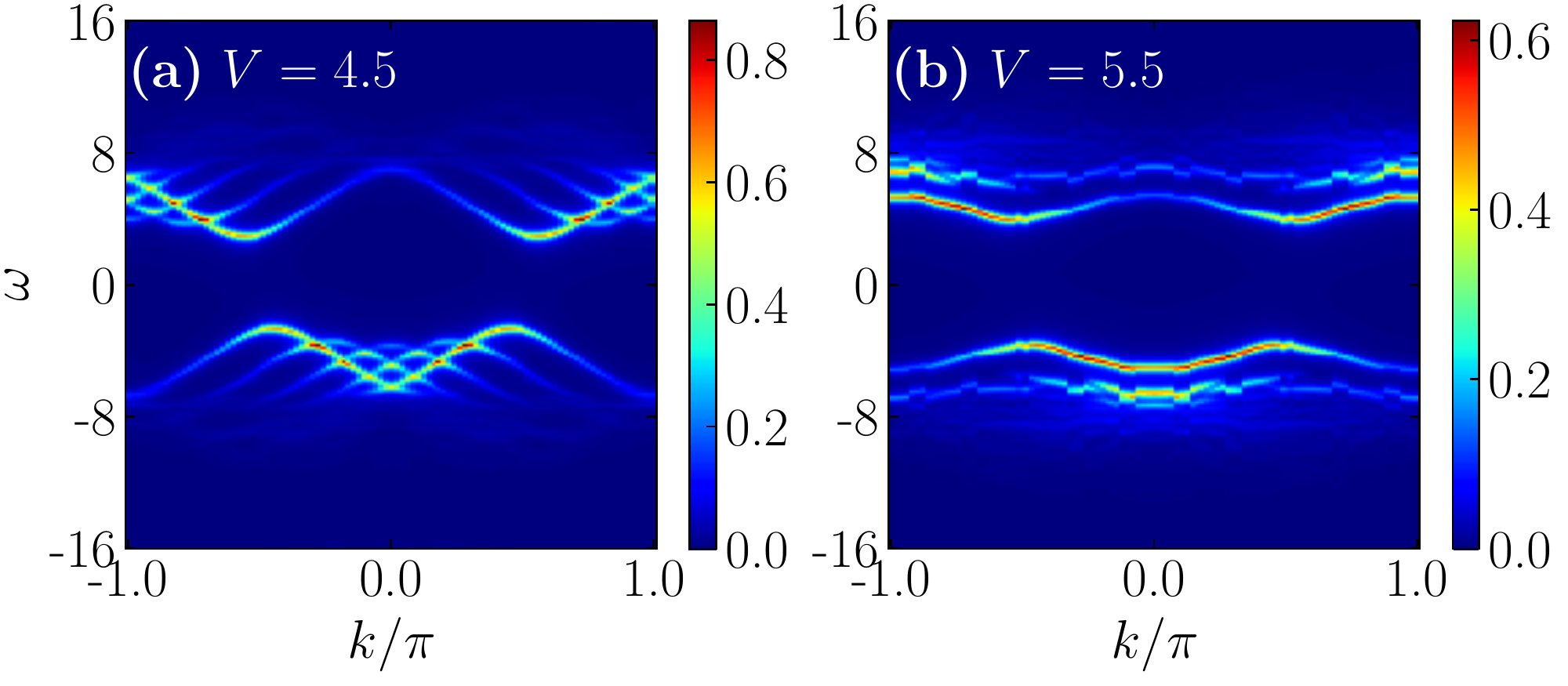}
\caption{(Color online) The equilibrium single-particle spectral function $I(k,\omega)$ for the half-filled EHM with $U=10.0$, and (a) $V=4.5$, (b) $V=5.5$, respectively.}
\label{fig_equilibrium}
\end{figure}

In the following discussions, we set the on-site repulsion $U=10.0$ so that the system is in the strong-coupling regime; the Hamiltonian before and after quench are denoted as $H_i$ and $H_f$, respectively. Throughout the main text, the lattice size is set to be $L = 10$. While in the Appendix the results for $L = 14$ are presented for comparison and for finite-size analysis. $I(k, \omega, t)$ is usually denoted as $I(k, \omega, \Delta t)$, with $\Delta t$ specifying the evolution time after quantum quenches.

\section{Results and discussions}\label{sec_pump}
\subsection{$I(k,\omega)$  before quench}\label{1sec_pump}

Before the discussion of the quench dynamics, let us first examine the single-particle spectral function $I(k,\omega)$ of the half-filled EHM in equilibrium.
Figures.~\ref{fig_equilibrium}(a) and \ref{fig_equilibrium}(b) show the results of systems with $V=4.5$ and $V=5.5$, respectively.
For the detailed results of $V=0.0$ and $V=7.0$, which are also essential in the present discussions, please consult Figs. 1(a) and 1(f) in Ref.~\onlinecite{Shao20}.
Remind that we set $U=10.0$ and the system is in the SDW (CDW) phase when $V\lesssim5.0$ ($V\gtrsim5.0$). In the $\omega$ space, a Lorentzian broadening of $0.2$ is introduced. The spectrum below (above) $\omega=0$ is exclusively composed of $I_{-}(k,\omega)$ [$I_{+}(k,\omega)$], which is known as the lower (upper) Hubbard band. In the SDW phase with $V=0.0$ and $=4.5$, the single-particle spectra include some interlaced ``stripes''. It is due to the finite-size effects and has been discussed in Refs.~\onlinecite{Kim96, Aichhorn04, Kim2006}. In contrast, two separated bands in the CDW phase can be observed both above and below the Fermi surface $\omega=0$ (see results of $V=5.5$ and $V=7.0$).

\begin{figure}[t]
\centering
\includegraphics[width=0.5\textwidth]{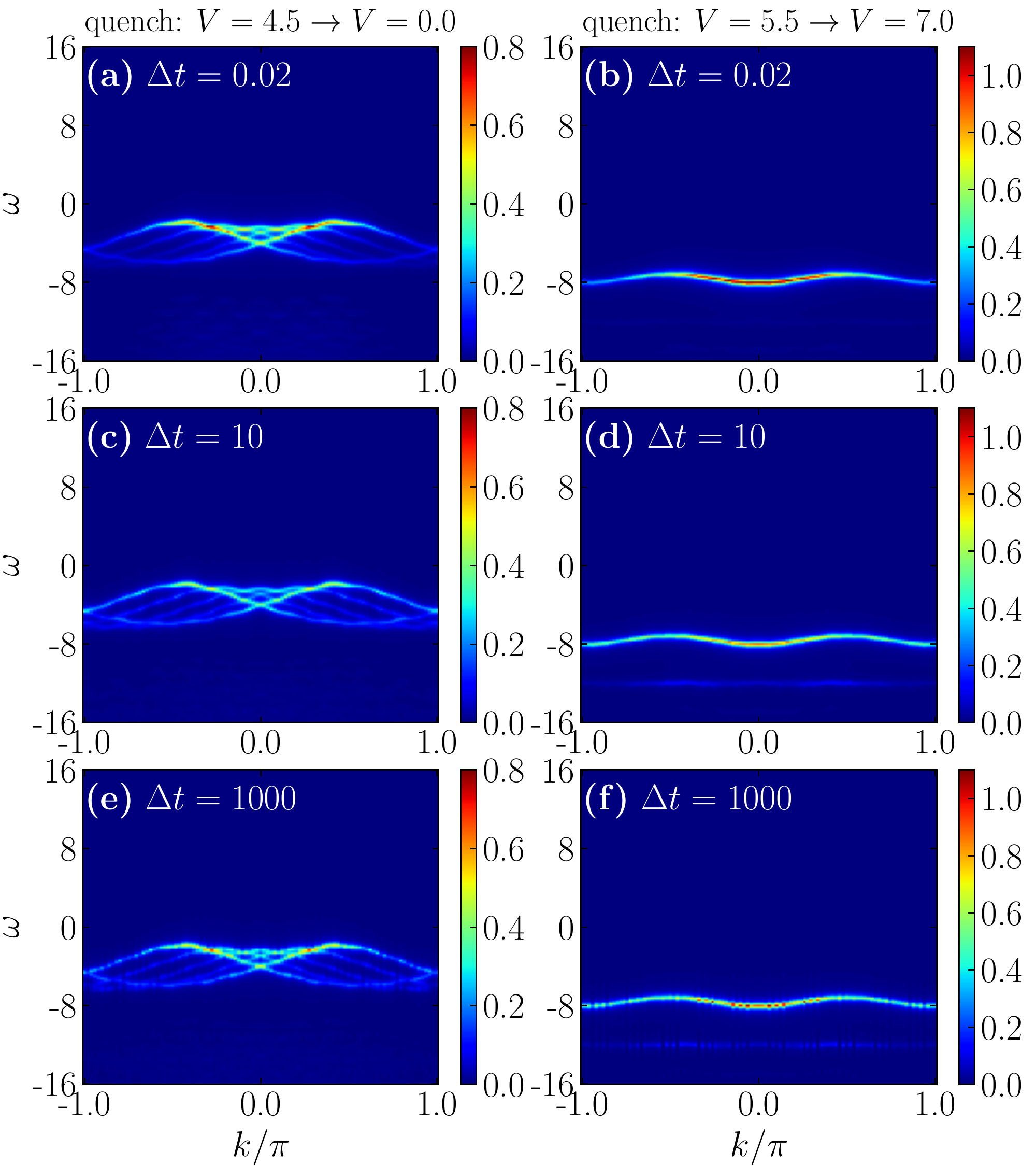}
\caption{(Color online) $I_{-}(k, \omega, \Delta t)$ of the half-filled EHM for $U=10.0$, and $V$ quenches from $4.5$ to $0.0$ in the left panel; $V$ quenches from $5.5$ to $7.0$ in the right panel. $\Delta t$ is the evolution time after quantum quench. For both quenches, we select $\Delta t=0.02$ in (a) and (b), $\Delta t=10$ in (c) and (d), $\Delta t=1000$ in (e) and (f), to present the snapshots of the time-dependent single-particle spectra.}
\label{fig_sameI}
\end{figure}

\begin{figure}[t]
\includegraphics[width=0.5\textwidth]{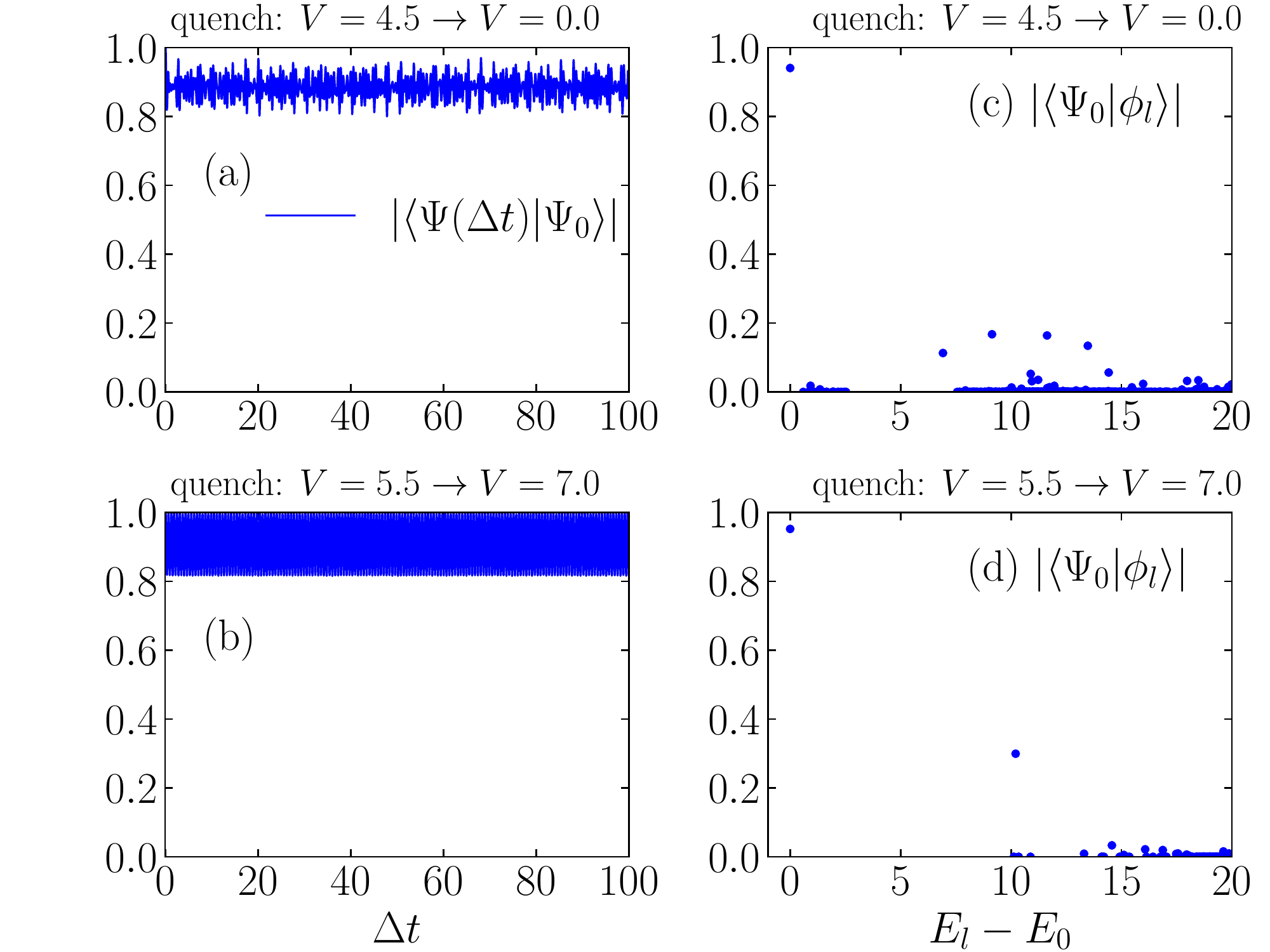}
\caption{
The overlaps of $\ket{\Psi(\Delta t)}$ with $\ket{\Psi_{0}}$ as a function of $\Delta t$ after (a) $V$ quench from $4.5$ to $0.0$, (b) $V$ quench from $5.5$ to $7.0$. $\ket{\Psi(\Delta{t}})$ and $\ket{\Psi_{0}}$ are the time-dependent wave function and the ground state of Hamiltonian before quench, respectively.
The overlaps between $\ket{\Psi_{0}}$ and $\ket{\phi_{l}}$ as a function of $E_{l}-E_{0}$ for the cases of (c) $V$ quench from $4.5$ to $0.0$ and (d) $V$ quench from $5.5$ to $7.0$. $\ket{\phi_{l}}$ is the $l$-th eigenstate of the quenched Hamiltonian in the $k=0$ momentum subspace, $E_{l}$ is the the eigenenergy of $\ket{\phi_{l}}$.}
\label{fig_sameO}
\end{figure}

\subsection{$V$ quench within the same phase}\label{1sec_same}

We move to the discussions of the non-equilibrium dynamics induced by a sudden interaction quench. The $V$-quench scenario can be categorized into two situations: the quenches within the same phase, and the ones across the critical point. In this subsection, let us focus on the first. In the left panel of Fig.~\ref{fig_sameI}, we show three snapshots of the time-dependent electron-removal spectral function $I_{-}(k,\omega,{\Delta}t)$ of the half-filled EHM after $V$ quench from $4.5$ to $0.0$ in the SDW phase. Note that in the remaining discussions, only the results of $I_{-}(k,\omega,{\Delta}t)$ are shown because $I_{+}(k,\omega,{\Delta}t)$ is always centrosymmetric to $I_{-}(k,\omega,{\Delta}t)$. We present the results of $\Delta t=0.02$, $10$ and $1000$ in Figs.~\ref{fig_sameI}(a), (c) and (e), respectively. Note that in our time-dependent Lanczos method, the minimum time interval has been set to be $\delta t=0.02$.
Even so, we find that just after the $V$ quench from $4.5$ to $0.0$, $I_{-}(k,\omega,{\Delta}t)$ becomes very close to the equilibrium result of $V=0.0$ [See Fig.~1(a) in Ref.~\onlinecite{Shao20}]. Subsequently, the basic spectral structure does not change with only slight spectral weight fluctuations taking place. Similar results can be found in the right panel of Fig.~\ref{fig_sameI}, where we show $I_{-}(k,\omega,{\Delta}t)$ after $V$ quench from $5.5$ to $7.0$ in the CDW for $\Delta t=0.02$, $10$ and $1000$, respectively. And we note that in a similar fashion, $I_{-}(k,\omega,{\Delta}t)$ is already close to the equilibrium result of $V=7.0$ [See Fig.~1 (f) in Ref.~\onlinecite{Shao20}] immediately after the quench. The phenomena can be understood as following.

In the quench dynamics, the initial state can be expanded in terms of the eigenstates of the quenched Hamiltonian $H_f$ (restricted to the $k = 0$ momentum subspace since the global interaction quench does not break translational symmetry) as $\ket{\Psi_{0}}=\sum^{}_{l}C_{l}|\phi_l\rangle$. Using $H_f|\phi_l\rangle=E_l|\phi_l\rangle$, we obtain $|{\Psi(\Delta{t})}\rangle\equiv \text{exp}(-iH_f\Delta t)\ket{\Psi_{0}}=\sum^{}_{l}C_{l}\text{exp}(-iE_l\Delta t)|\phi_l\rangle$. The overlap between time-dependent wave function $\ket{\Psi({\Delta}t)}$ and the initial state $\ket{\Psi_{0}}$ can be written as
\begin{eqnarray}
|\braket{\Psi_{0}}{\Psi(\Delta{t})}|&=&|\sum_{l}C^{*}_{l}C_{l}e^{-\mathrm{i}E_{l}\Delta t}|.
\label{12}
\end{eqnarray}
We can find that the time evolution of $|\braket{\Psi_{0}}{\Psi(\Delta{t})}|$ depends completely on the sets of $\{C_l\}$ and $\{E_l\}$. We show the results of $|\braket{\Psi_{0}}{\Psi(\Delta{t})}|$ in Figs.~\ref{fig_sameO}(a) and \ref{fig_sameO}(b) for the cases of $V$ quench from $4.5$ to $0.0$ and $V$ quench from $5.5$ to $7.0$, respectively. The time-resolved overlaps in both are close to $1$ due to the dominating $C_0$ in all $C_l$'s, which means that the initial state overlaps largely with the ground state of the quenched Hamiltonian. This can be seen in Figs.~\ref{fig_sameO}(c) and \ref{fig_sameO}(d), where we show the distribution of $|C_{l}|$ (i.e., $|\langle\Psi_0|\phi_l\rangle|$) as a function of $E_{l}-E_{0}$ for the aforesaid two quench cases. It is known that at the first-order critical point, the wave functions and the order parameters of the two phases (SDW and CDW for our study) abruptly change, while the ground states within the same phase keep their similarity. This is why quenching a Hamiltonian within the same phase usually produces a trivial time evolution.

\begin{figure}[t]
\includegraphics[width=0.5\textwidth]{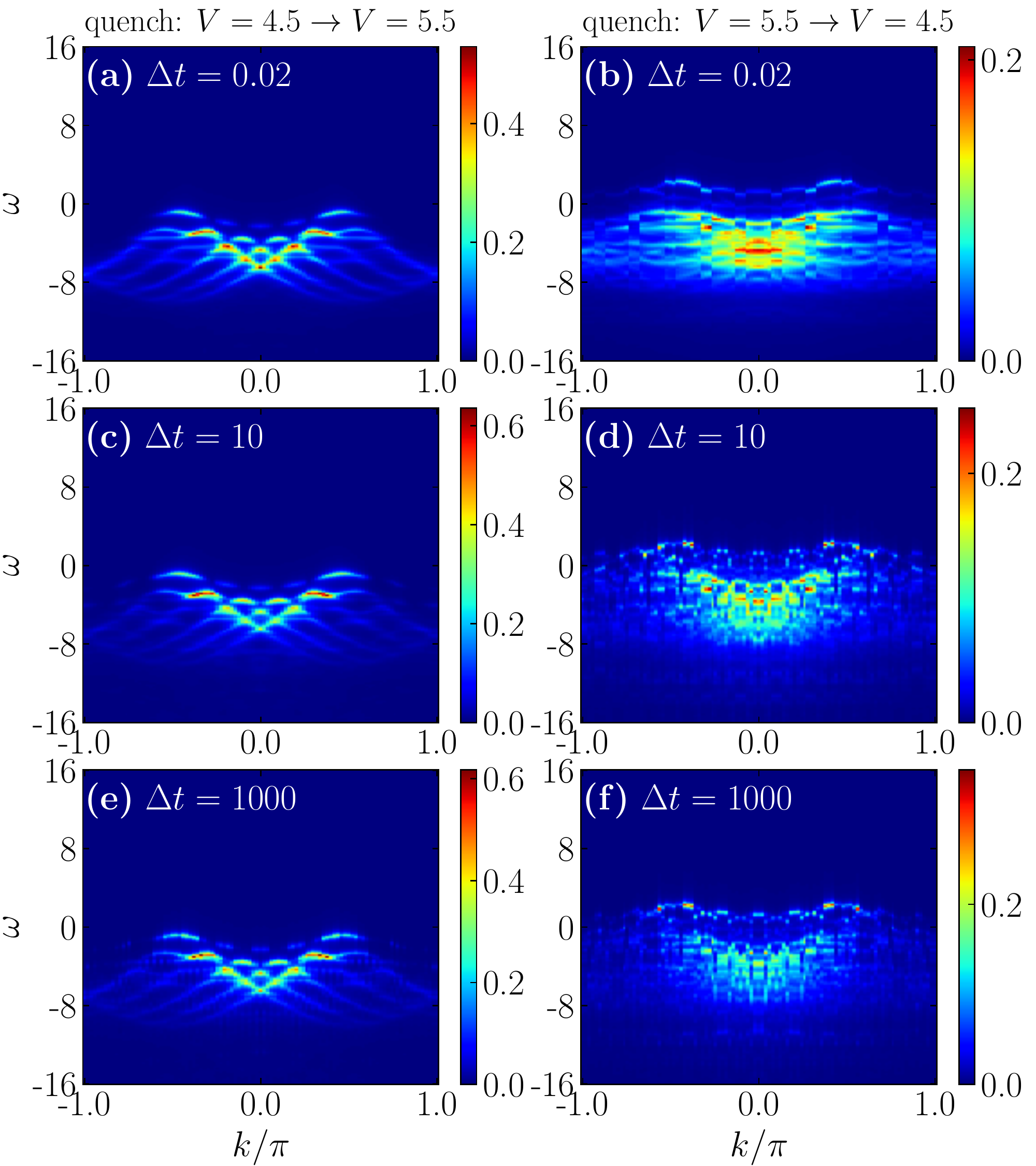}
\caption{(Color online)$I_{-}(k, \omega, \Delta t)$ of the half-filled EHM for $U=10.0$, and $V$ quenches from $4.5$ to $5.5$ in the left panel; $V$ quenches from $5.5$ to $4.5$ in the right panel. $\Delta t$ is the evolution time after quantum quench. For both quenches, we select $\Delta t=0.02$ in (a) and (b), $\Delta t=10$ in (c) and (d), $\Delta t=1000$ in (e) and (f), to present the snapshots of the time-dependent single-particle spectra.}
\label{fig_4}
\end{figure}

In addition, for the time-dependent single-particle spectral function
\begin{eqnarray}
&I_{-}(k,\omega,\Delta t)=  \\
&-\frac{1}{\pi}\text{Im}\left(\langle\Psi^{\kappa}(\Delta t)|c_{k_0,\sigma}^{\dag}\frac{1}{\omega-(H_f^{\kappa}-E^{\kappa}(\Delta t))}c_{k_0,\sigma}|\Psi^{\kappa}(\Delta t)\rangle\right), \nonumber
\label{I-}
\end{eqnarray}
it is determined by the quenched Hamiltonian $H_f^{\kappa}$ and the time-evolving wave function $|\Psi^{\kappa}(\Delta t)\rangle$. We speculate that if $|\Psi^{\kappa}(\Delta t=0)\rangle$ (i.e., $|\Psi_0^{\kappa}\rangle$) has a large overlap with the ground state of the quenched Hamiltonian, $I_{-}(k,\omega,\Delta t)$ can evolve quickly to the corresponding equilibrium values of the quenched system.

\begin{figure}[t]
\includegraphics[width=0.5\textwidth]{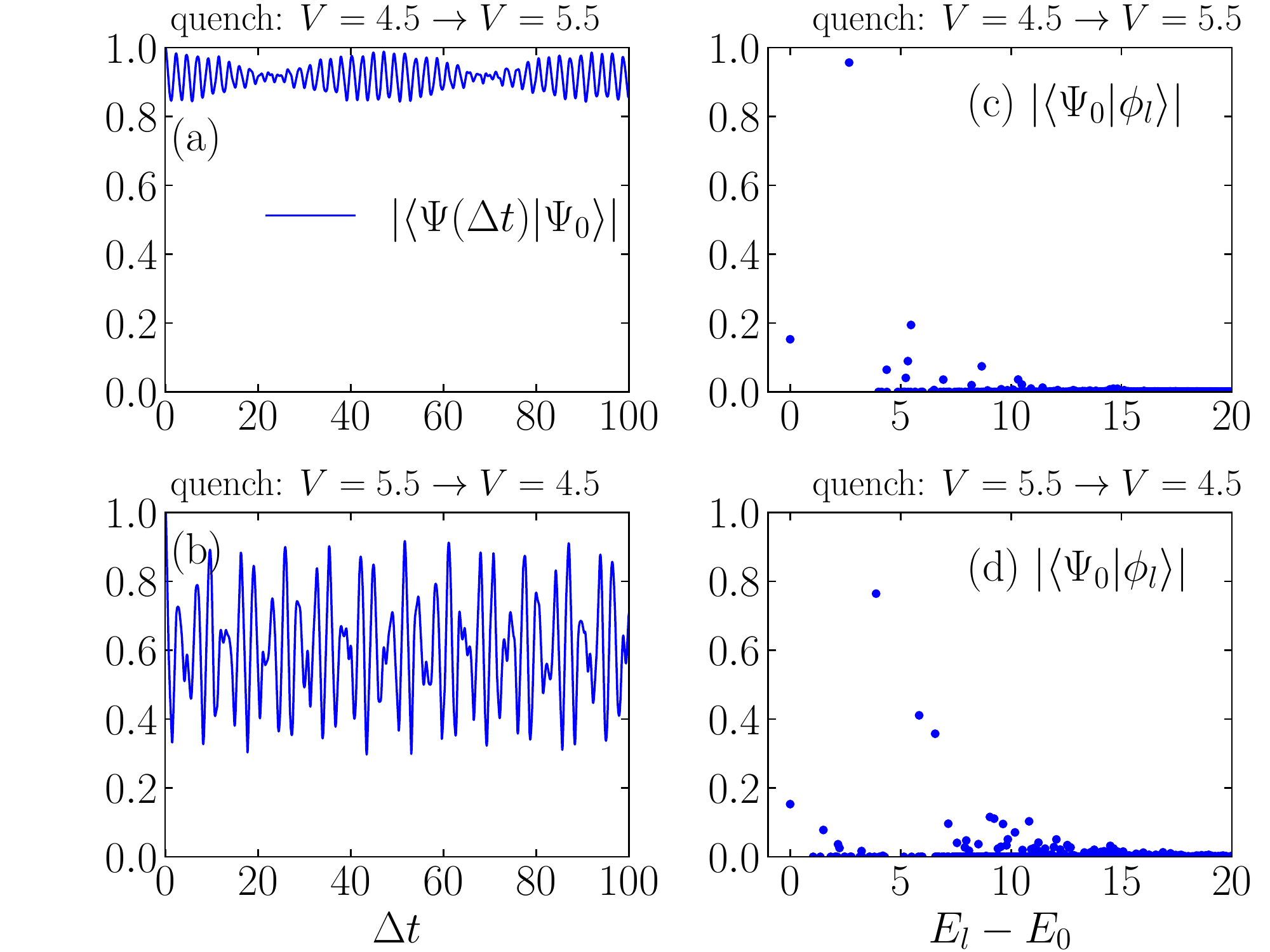}
\caption{
The overlaps of $\ket{\Psi(\Delta t)}$ with $\ket{\Psi_{0}}$ as a function of $\Delta t$ after (a) $V$ quench from $4.5$ to $5.5$, (b) $V$ quench from $5.5$ to $4.5$. $\ket{\Psi(\Delta{t}})$ and $\ket{\Psi_{0}}$ are the time-dependent wave function and the ground state of Hamiltonian before quench, respectively.
The overlaps between $\ket{\Psi_{0}}$ and $\ket{\phi_{l}}$ as a function of $E_{l}-E_{0}$ for the cases of (c) $V$ quench from $4.5$ to $5.5$ and (d) $V$ quench from $5.5$ to $4.5$. $\ket{\phi_{l}}$ is the $l$-th eigenstate of the quenched Hamiltonian in the $k=0$ momentum subspace, $E_{l}$ is the the corresponding eigenenergy of $\ket{\phi_{l}}$.}
\label{fig_5}
\end{figure}

\subsection{$V$ quench across the critical point}\label{1sec_different}

Different from the previous cases of quenches within the same phase, three snapshots of $I_{-}(k, \omega, \Delta t)$ for $V$ quench from $4.5$ to $5.5$ ($V$ quench from $5.5$ to $4.5$) are shown in the left (right) panels of Fig.~\ref{fig_4}. For both cases, the quench crosses the critical point, $V_{C}\approx 5.0$. $\Delta t=0.02$, $10$ and $1000$ are also chosen in Figs.~\ref{fig_4}(a) and \ref{fig_4}(b), \ref{fig_4}(c) and \ref{fig_4}(d), as well as \ref{fig_4}(e) and \ref{fig_4}(f), respectively. Compared to Fig.~\ref{fig_equilibrium}(a), we find that $I_{-}(k, \omega, \Delta t)$ for $V$ quench from $4.5$ to $5.5$ keeps significant memories of the initial state for a long time even though some spectral weight has shifted towards the high-energy part. It can be still understood by analyzing the distribution of the initial state among the eigenstates of $H_f$. The overlaps $|C_{l}|$ (i.e., $|\langle\Psi_0|\phi_l\rangle|$) as a function of $E_{l}-E_{0}$ are shown in Fig.~\ref{fig_5}(c) for the case of $V$ quench from $4.5$ to $5.5$. Here $C_{1}$ (the overlap of the initial state with the first excitation state of the quenched Hamiltonian) instead of $C_0$ distinctly dominates, which guarantees a sufficient overlap of $\ket{\Psi(\Delta t)}$ with the initial state $\ket{\Psi_0}$ during the evolution time we have measured [in terms of Eq.~(\ref{12})], as shown in Fig.~\ref{fig_5}(a). As a consequence, no significant change of the time-dependent single-particle spectra after quench has been observed.

On the other hand, the spectra after quench from $5.5$ to $4.5$ have little similarity to the equilibrium results either for $V=5.5$ or for $V=4.5$, as shown in the right panels of Fig.~\ref{fig_4}.  The reason can be read from Fig.~\ref{fig_5}(d), where three considerable components of $C_l$ are spotted, with one close to $0.8$ and the other two around $0.4$.
As a result of the interference between these eigenmodes, the time-dependent overlap $|\langle\Psi(\Delta t)|\Psi_0\rangle|$ in Fig.~\ref{fig_5}(b) oscillates with significant amplitude.
Correspondingly, the time-dependent spectra become more evenly distributed, as shown in the right panels of Fig.~\ref{fig_4}.

In Ref.~\onlinecite{Jeckelmann03}, four types of excitations are found in the Mott insulating phase (the SDW phase in our case) with each one dominating the low-energy spectrum in a particular region of the parameter space. The proliferation of low-lying excitations in the spectrum of the SDW side may produce a diverse distribution of the CDW ground state among the quenched SDW Hamiltonian eigenmodes [as shown in Fig.~\ref{fig_5}(d)].
The recognition of such difference can provide a plausible guide to understanding the asymmetry between the phase-crossing quenches with opposite directions.

\begin{figure}[t]
\centering
\includegraphics[width=0.5\textwidth]{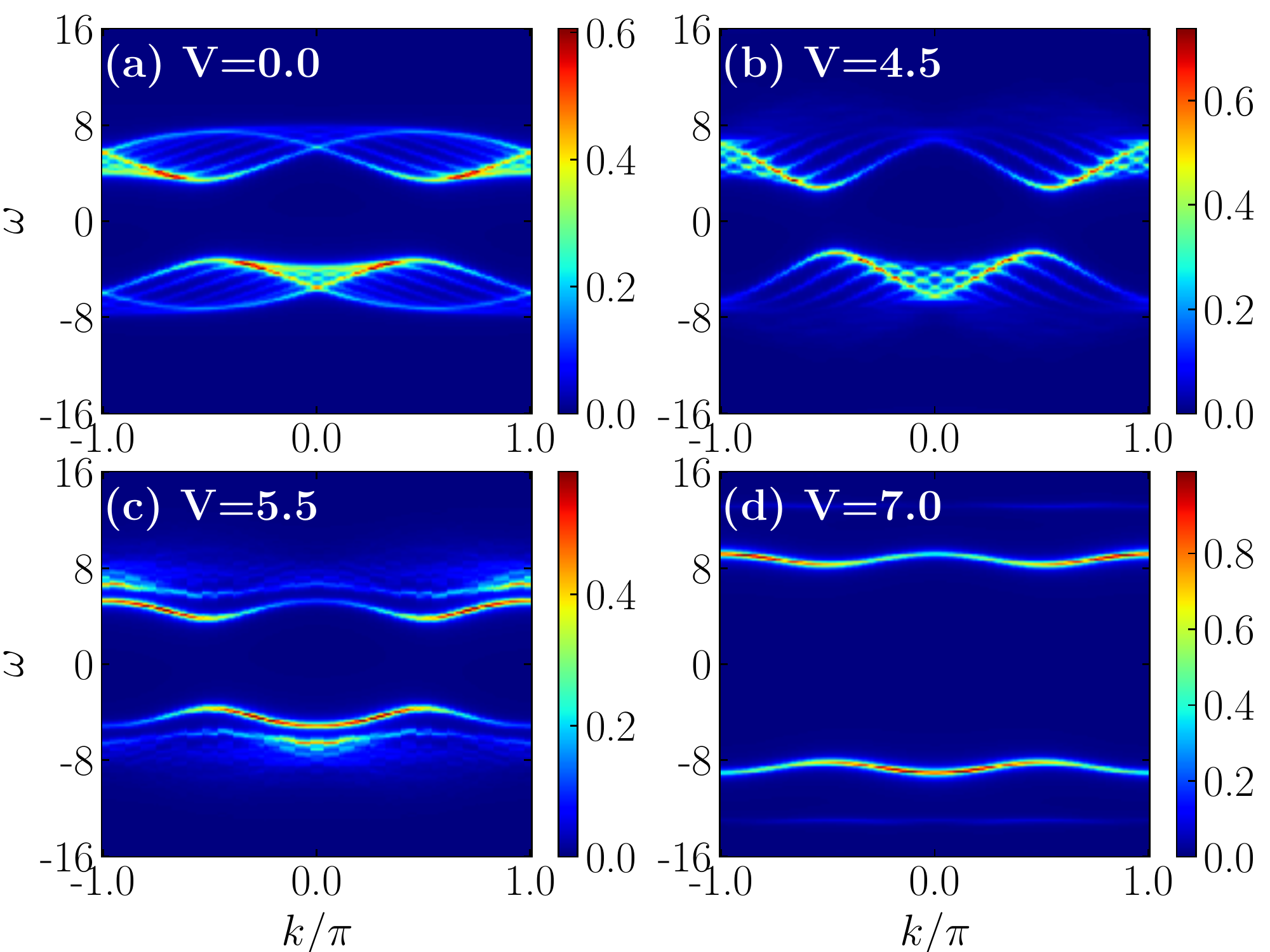}
\caption{(Color online) The equilibrium single-particle spectral function $I(k,\omega)$ for the half-filled EHM with (a) $V=0.0$, (b) $V=4.5$, (c) $V=5.5$ and (d) $V=7.0$ respectively. Other parameters: $L=14$ and $U=10.0$.}
\label{fig_A1}
\end{figure}

\section{Conclusion}\label{sec_conclusion}

To conclude, we studied the quench dynamics of the one-dimensional extended Hubbard model at half-filling.
In strong-coupling regime, the ground states of this model with small and large nearest-neighbor interaction $V$ are in the spin-density-wave and charge-density-wave phases, respectively.
We found that if the quench happens within the same phase, there is a considerable overlap between the initial state $|\Psi_0\rangle$ and the ground state of $H_f$ so that the spectrum evolves quickly towards the ground-state counterpart of the quenched Hamiltonian. If the quench crosses the critical point, however, the evolution of the time-dependent wave function and the single-particle spectrum mainly depend on the overlaps of $|\Psi_0\rangle$ with the eigenstates of $H_f$. Our findings could offer the possibility to detect the phase transition in ultracold atom systems.

\begin{acknowledgments}
C. S. acknowledges support from the National Natural Science Foundation of China (NSFC; Grant No. 12104229) and the Fundamental Research Funds for the Central Universities (Grant No. 30922010803).
R. F. acknowledges supports from NSFC (Grants No. 11974185) and the Natural Science Foundation of Jiangsu Province (Grant No. BK20170032).
H. L. acknowledges support from NSFC (Grants No. 11874187, No. 12174168 and No. 12047501).
\end{acknowledgments}

\begin{figure}[t]
\centering
\includegraphics[width=0.5\textwidth]{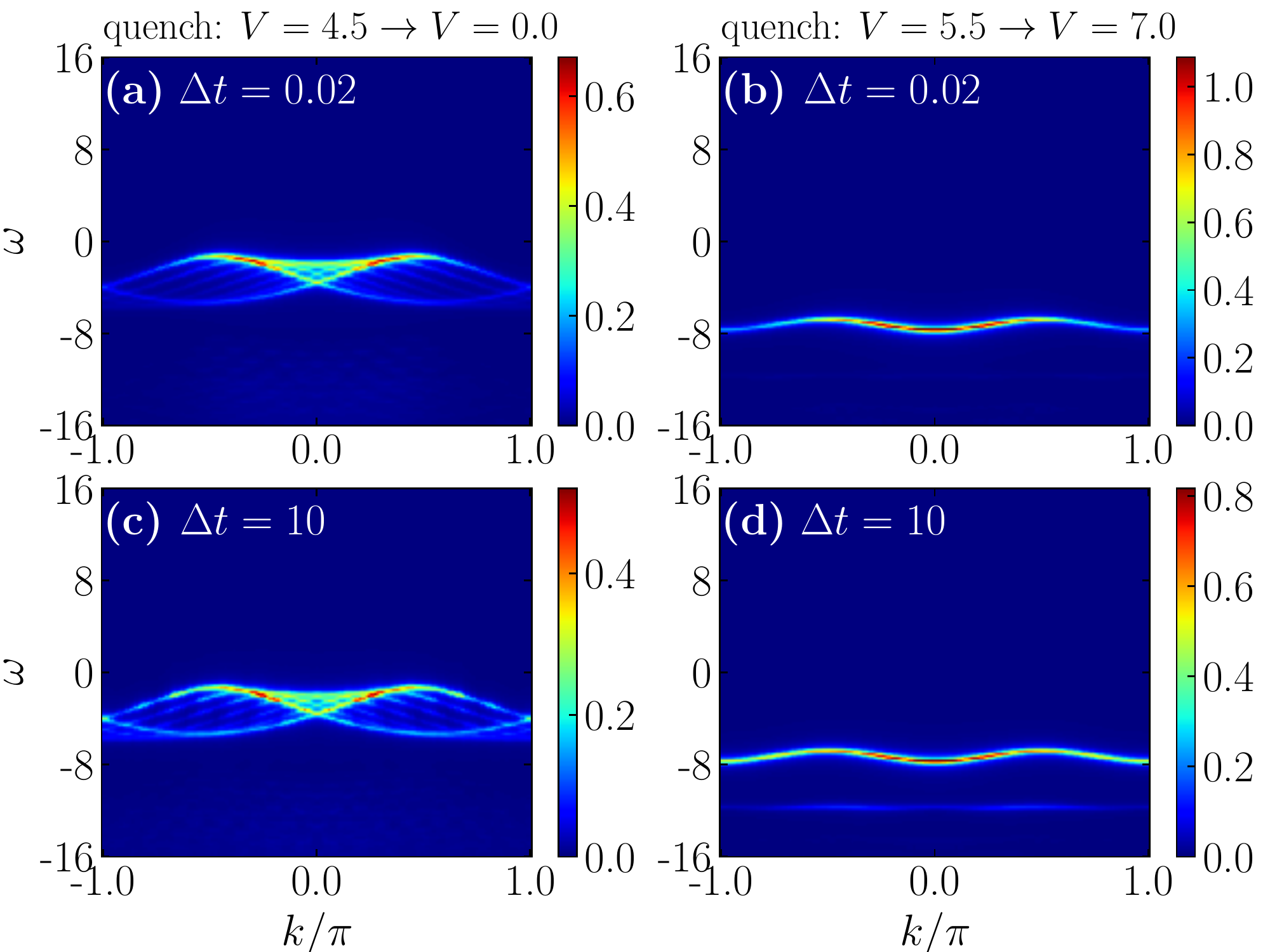}
\caption{(Color online) $I_{-}(k, \omega, \Delta t)$ of the half-filled EHM for $V$ quenches from $4.5$ to $0.0$ in the left panel; $V$ quenches from $5.5$ to $7.0$ in the right panel. $\Delta t$ is the evolution time after quantum quench. For both quenches, we select $\Delta t=0.02$ in (a) and (b), $\Delta t=10$ in (c) and (d), to present the snapshots of the time-dependent single-particle spectra. Other parameters: $L=14$ and $U=10.0$.}
\label{fig_A2}
\end{figure}

\appendix*
\section{The Finite-Size Analysis}\label{Appendix}

In the main text, we have chosen the lattice size $L=10$ for the demonstration.  In this Appendix, we present the single-particle spectral
function in and out of equilibrium with lattice size $L=14$. For the half-filled EHM with $U=10.0$, the equilibrium single-particle spectra with $V=0.0$, $V=4.5$, $V=5.5$ and $V=7.0$ are shown in Figs.~\ref{fig_A1}(a), \ref{fig_A1}(b), \ref{fig_A1}(c) and \ref{fig_A1}(d), respectively. In the SDW phase with $V=0.0$ and $V=4.5$, there are more interlaced ``stripes" compared to the results with lattice size $L=10$. As we mentioned before, this is due to the finite-size effect and the spectrum will split into a spinon and a holon band in the thermodynamic limit~\cite{Kim96, Aichhorn04, Kim2006}. In the CDW phase with $V=5.5$ and $V=7.0$, the spectra are nearly same to the results with $L=10$, where two separated flat bands can be observed both above and below the Fermi surface $\omega=0$.

For $V$ quenches within the same phase, we show in Fig.~\ref{fig_A2} two snapshots ($\Delta t=0.02$ and $10$) of the time-dependent electron-removal spectral function $I_{-}(k,\omega,{\Delta}t)$ after $V$ quench from $4.5$ to $0.0$ in the left panel and $V$ quench from $5.5$ to $7.0$ in the right panel, respectively. Similar to that with $L=10$, $I_{-}(k,\omega,{\Delta}t)$ becomes very close to the equilibrium result of the quenched Hamiltonian just after quench. Unfortunately, we can not provide the overlap distribution of the initial state among the eigenstates of the quenched Hamiltonian because the full ED calculation of the $L=14$ system is unreachable for the current computational resources. However, the basic conclusions on the time-dependent spectral function remain unchange.

\begin{figure}[t]
\centering
\includegraphics[width=0.5\textwidth]{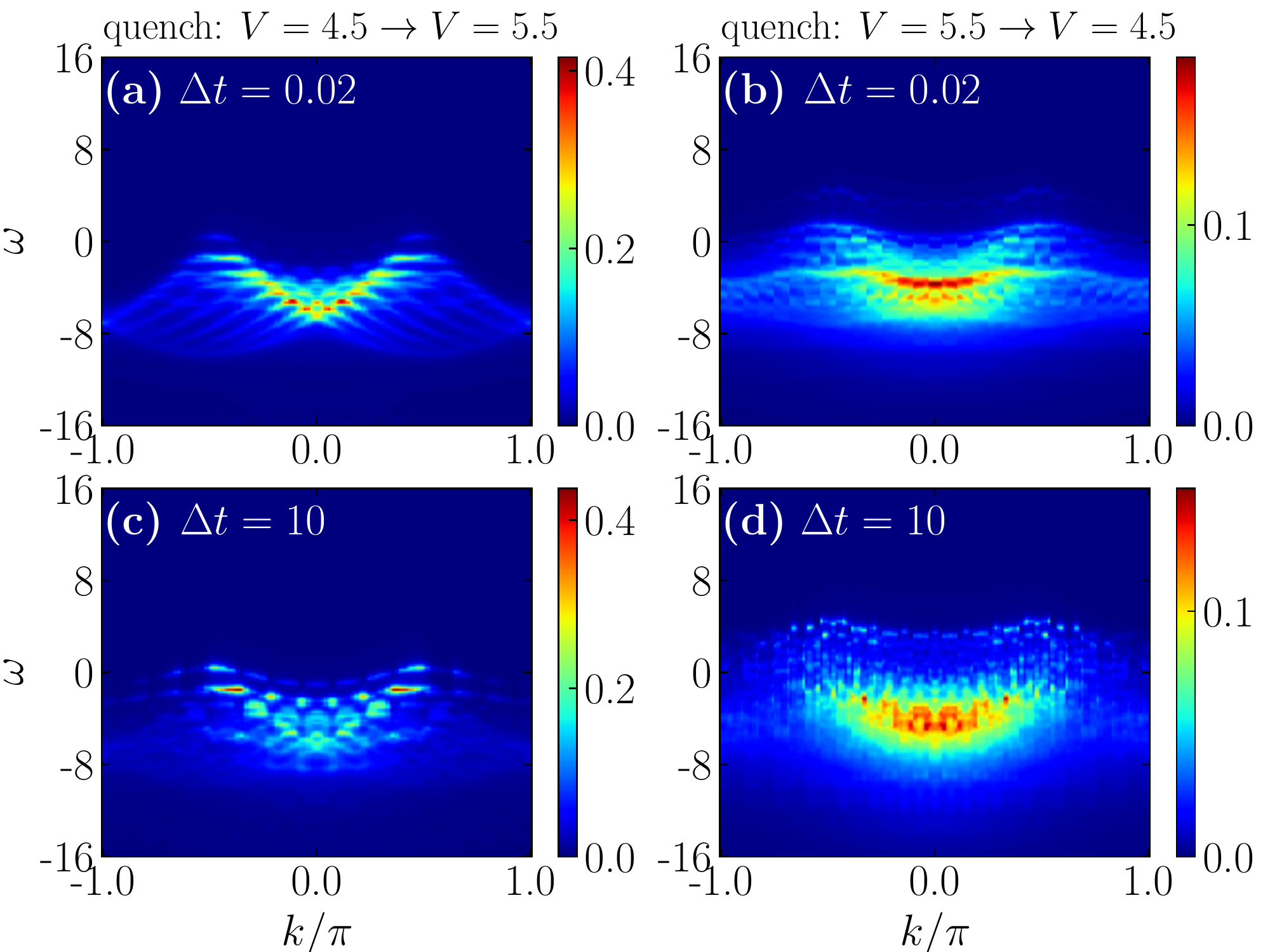}
\caption{(Color online) $I_{-}(k, \omega, \Delta t)$ of the half-filled EHM for $V$ quenches from $4.5$ to $5.5$ in the left panel; $V$ quenches from $5.5$ to $4.5$ in the right panel. $\Delta t$ is the evolution time after quantum quench. For both quenches, we select $\Delta t=0.02$ in (a) and (b), $\Delta t=10$ in (c) and (d), to present the snapshots of the time-dependent single-particle spectra. Other parameters: $L=14$ and $U=10.0$.}
\label{fig_A3}
\end{figure}

For $V$ quench from $4.5$ to $5.5$ across the critical point, $I_{-}(k, \omega, \Delta t)$ still keeps some memories of the initial state with more spectral weight shifting to the high-energy part, as shown in Figs.~\ref{fig_A3}(a) and \ref{fig_A3}(c). On the other hand, $I_{-}(k, \omega, \Delta t)$ after quench from $5.5$ to $4.5$ have little similarity to the equilibrium spectra either for $V=5.5$ or $V=4.5$, as shown in Figs.~\ref{fig_A3}(b) and \ref{fig_A3}(d), respectively. All these features consistent with the results of $L=10$ in the main text.


\begin{thebibliography}{50}%
\makeatletter
\providecommand \@ifxundefined [1]{%
 \@ifx{#1\undefined}
}%
\providecommand \@ifnum [1]{%
 \ifnum #1\expandafter \@firstoftwo
 \else \expandafter \@secondoftwo
 \fi
}%
\providecommand \@ifx [1]{%
 \ifx #1\expandafter \@firstoftwo
 \else \expandafter \@secondoftwo
 \fi
}%
\providecommand \natexlab [1]{#1}%
\providecommand \enquote  [1]{``#1''}%
\providecommand \bibnamefont  [1]{#1}%
\providecommand \bibfnamefont [1]{#1}%
\providecommand \citenamefont [1]{#1}%
\providecommand \href@noop [0]{\@secondoftwo}%
\providecommand \href [0]{\begingroup \@sanitize@url \@href}%
\providecommand \@href[1]{\@@startlink{#1}\@@href}%
\providecommand \@@href[1]{\endgroup#1\@@endlink}%
\providecommand \@sanitize@url [0]{\catcode `\\12\catcode `\$12\catcode
  `\&12\catcode `\#12\catcode `\^12\catcode `\_12\catcode `\%12\relax}%
\providecommand \@@startlink[1]{}%
\providecommand \@@endlink[0]{}%
\providecommand \url  [0]{\begingroup\@sanitize@url \@url }%
\providecommand \@url [1]{\endgroup\@href {#1}{\urlprefix }}%
\providecommand \urlprefix  [0]{URL }%
\providecommand \Eprint [0]{\href }%
\providecommand \doibase [0]{http://dx.doi.org/}%
\providecommand \selectlanguage [0]{\@gobble}%
\providecommand \bibinfo  [0]{\@secondoftwo}%
\providecommand \bibfield  [0]{\@secondoftwo}%
\providecommand \translation [1]{[#1]}%
\providecommand \BibitemOpen [0]{}%
\providecommand \bibitemStop [0]{}%
\providecommand \bibitemNoStop [0]{.\EOS\space}%
\providecommand \EOS [0]{\spacefactor3000\relax}%
\providecommand \BibitemShut  [1]{\csname bibitem#1\endcsname}%
\let\auto@bib@innerbib\@empty
\bibitem [{\citenamefont {Polkovnikov}\ \emph {et~al.}(2011)\citenamefont
  {Polkovnikov}, \citenamefont {Sengupta}, \citenamefont {Silva},\ and\
  \citenamefont {Vengalattore}}]{Polkovnikov11}%
  \BibitemOpen
  \bibfield  {author} {\bibinfo {author} {\bibfnamefont {A.}~\bibnamefont
  {Polkovnikov}}, \bibinfo {author} {\bibfnamefont {K.}~\bibnamefont
  {Sengupta}}, \bibinfo {author} {\bibfnamefont {A.}~\bibnamefont {Silva}}, \
  and\ \bibinfo {author} {\bibfnamefont {M.}~\bibnamefont {Vengalattore}},\
  }\href {\doibase 10.1103/RevModPhys.83.863} {\bibfield  {journal} {\bibinfo
  {journal} {Rev. Mod. Phys.}\ }\textbf {\bibinfo {volume} {83}},\ \bibinfo
  {pages} {863} (\bibinfo {year} {2011})}\BibitemShut {NoStop}%
\bibitem [{\citenamefont {Dziarmaga}(2010)}]{Jacek10}%
  \BibitemOpen
  \bibfield  {author} {\bibinfo {author} {\bibfnamefont {J.}~\bibnamefont
  {Dziarmaga}},\ }\href {\doibase 10.1080/00018732.2010.514702} {\bibfield
  {journal} {\bibinfo  {journal} {Advances in Physics}\ }\textbf {\bibinfo
  {volume} {59}},\ \bibinfo {pages} {1063} (\bibinfo {year}
  {2010})}\BibitemShut {NoStop}%
\bibitem [{\citenamefont {Cazalilla}\ and\ \citenamefont
  {Rigol}(2010)}]{Cazalilla_2010}%
  \BibitemOpen
  \bibfield  {author} {\bibinfo {author} {\bibfnamefont {M.~A.}\ \bibnamefont
  {Cazalilla}}\ and\ \bibinfo {author} {\bibfnamefont {M.}~\bibnamefont
  {Rigol}},\ }\href {\doibase 10.1088/1367-2630/12/5/055006} {\bibfield
  {journal} {\bibinfo  {journal} {New Journal of Physics}\ }\textbf {\bibinfo
  {volume} {12}},\ \bibinfo {pages} {055006} (\bibinfo {year}
  {2010})}\BibitemShut {NoStop}%
\bibitem [{\citenamefont {Kinoshita}\ \emph {et~al.}(2006)\citenamefont
  {Kinoshita}, \citenamefont {Wenger},\ and\ \citenamefont
  {Weiss}}]{Kinoshita2006}%
  \BibitemOpen
  \bibfield  {author} {\bibinfo {author} {\bibfnamefont {T.}~\bibnamefont
  {Kinoshita}}, \bibinfo {author} {\bibfnamefont {T.}~\bibnamefont {Wenger}}, \
  and\ \bibinfo {author} {\bibfnamefont {D.~S.}\ \bibnamefont {Weiss}},\ }\href
  {\doibase 10.1038/nature04693} {\bibfield  {journal} {\bibinfo  {journal}
  {Nature}\ }\textbf {\bibinfo {volume} {440}},\ \bibinfo {pages} {900}
  (\bibinfo {year} {2006})}\BibitemShut {NoStop}%
\bibitem [{\citenamefont {Gring}\ \emph {et~al.}(2012)\citenamefont {Gring},
  \citenamefont {Kuhnert}, \citenamefont {Langen}, \citenamefont {Kitagawa},
  \citenamefont {Rauer}, \citenamefont {Schreitl}, \citenamefont {Mazets},
  \citenamefont {Smith}, \citenamefont {Demler},\ and\ \citenamefont
  {Schmiedmayer}}]{Gring12}%
  \BibitemOpen
  \bibfield  {author} {\bibinfo {author} {\bibfnamefont {M.}~\bibnamefont
  {Gring}}, \bibinfo {author} {\bibfnamefont {M.}~\bibnamefont {Kuhnert}},
  \bibinfo {author} {\bibfnamefont {T.}~\bibnamefont {Langen}}, \bibinfo
  {author} {\bibfnamefont {T.}~\bibnamefont {Kitagawa}}, \bibinfo {author}
  {\bibfnamefont {B.}~\bibnamefont {Rauer}}, \bibinfo {author} {\bibfnamefont
  {M.}~\bibnamefont {Schreitl}}, \bibinfo {author} {\bibfnamefont
  {I.}~\bibnamefont {Mazets}}, \bibinfo {author} {\bibfnamefont {D.~A.}\
  \bibnamefont {Smith}}, \bibinfo {author} {\bibfnamefont {E.}~\bibnamefont
  {Demler}}, \ and\ \bibinfo {author} {\bibfnamefont {J.}~\bibnamefont
  {Schmiedmayer}},\ }\href {\doibase 10.1126/science.1224953} {\bibfield
  {journal} {\bibinfo  {journal} {Science}\ }\textbf {\bibinfo {volume}
  {337}},\ \bibinfo {pages} {1318} (\bibinfo {year} {2012})}\BibitemShut
  {NoStop}%
\bibitem [{\citenamefont {Trotzky}\ \emph {et~al.}(2012)\citenamefont
  {Trotzky}, \citenamefont {Chen}, \citenamefont {Flesch}, \citenamefont
  {McCulloch}, \citenamefont {Schollw{\"{o}}ck}, \citenamefont {Eisert},\ and\
  \citenamefont {Bloch}}]{Trotzky2012}%
  \BibitemOpen
  \bibfield  {author} {\bibinfo {author} {\bibfnamefont {S.}~\bibnamefont
  {Trotzky}}, \bibinfo {author} {\bibfnamefont {Y.-A.}\ \bibnamefont {Chen}},
  \bibinfo {author} {\bibfnamefont {A.}~\bibnamefont {Flesch}}, \bibinfo
  {author} {\bibfnamefont {I.~P.}\ \bibnamefont {McCulloch}}, \bibinfo {author}
  {\bibfnamefont {U.}~\bibnamefont {Schollw{\"{o}}ck}}, \bibinfo {author}
  {\bibfnamefont {J.}~\bibnamefont {Eisert}}, \ and\ \bibinfo {author}
  {\bibfnamefont {I.}~\bibnamefont {Bloch}},\ }\href {\doibase
  10.1038/nphys2232} {\bibfield  {journal} {\bibinfo  {journal} {Nature
  Physics}\ }\textbf {\bibinfo {volume} {8}},\ \bibinfo {pages} {325} (\bibinfo
  {year} {2012})}\BibitemShut {NoStop}%
\bibitem [{\citenamefont {Rigol}\ \emph {et~al.}(2008)\citenamefont {Rigol},
  \citenamefont {Dunjko},\ and\ \citenamefont {Olshanii}}]{Rigol2008}%
  \BibitemOpen
  \bibfield  {author} {\bibinfo {author} {\bibfnamefont {M.}~\bibnamefont
  {Rigol}}, \bibinfo {author} {\bibfnamefont {V.}~\bibnamefont {Dunjko}}, \
  and\ \bibinfo {author} {\bibfnamefont {M.}~\bibnamefont {Olshanii}},\ }\href
  {\doibase 10.1038/nature06838} {\bibfield  {journal} {\bibinfo  {journal}
  {Nature}\ }\textbf {\bibinfo {volume} {452}},\ \bibinfo {pages} {854}
  (\bibinfo {year} {2008})}\BibitemShut {NoStop}%
\bibitem [{\citenamefont {D'Alessio}\ \emph {et~al.}(2016)\citenamefont
  {D'Alessio}, \citenamefont {Kafri}, \citenamefont {Polkovnikov},\ and\
  \citenamefont {Rigol}}]{Luca16}%
  \BibitemOpen
  \bibfield  {author} {\bibinfo {author} {\bibfnamefont {L.}~\bibnamefont
  {D'Alessio}}, \bibinfo {author} {\bibfnamefont {Y.}~\bibnamefont {Kafri}},
  \bibinfo {author} {\bibfnamefont {A.}~\bibnamefont {Polkovnikov}}, \ and\
  \bibinfo {author} {\bibfnamefont {M.}~\bibnamefont {Rigol}},\ }\href
  {\doibase 10.1080/00018732.2016.1198134} {\bibfield  {journal} {\bibinfo
  {journal} {Advances in Physics}\ }\textbf {\bibinfo {volume} {65}},\ \bibinfo
  {pages} {239} (\bibinfo {year} {2016})}\BibitemShut {NoStop}%
\bibitem [{\citenamefont {Deutsch}(2018)}]{Deutsch_2018}%
  \BibitemOpen
  \bibfield  {author} {\bibinfo {author} {\bibfnamefont {J.~M.}\ \bibnamefont
  {Deutsch}},\ }\href {\doibase 10.1088/1361-6633/aac9f1} {\bibfield  {journal}
  {\bibinfo  {journal} {Reports on Progress in Physics}\ }\textbf {\bibinfo
  {volume} {81}},\ \bibinfo {pages} {082001} (\bibinfo {year}
  {2018})}\BibitemShut {NoStop}%
\bibitem [{\citenamefont {Calabrese}(2020)}]{Calabrese2020}%
  \BibitemOpen
  \bibfield  {author} {\bibinfo {author} {\bibfnamefont {P.}~\bibnamefont
  {Calabrese}},\ }\href {\doibase 10.21468/SciPostPhysLectNotes.20} {\bibfield
  {journal} {\bibinfo  {journal} {SciPost Phys. Lect. Notes}\ ,\ \bibinfo
  {pages} {20}} (\bibinfo {year} {2020})}\BibitemShut {NoStop}%
\bibitem [{\citenamefont {Rigol}\ \emph {et~al.}(2007)\citenamefont {Rigol},
  \citenamefont {Dunjko}, \citenamefont {Yurovsky},\ and\ \citenamefont
  {Olshanii}}]{Rigol07}%
  \BibitemOpen
  \bibfield  {author} {\bibinfo {author} {\bibfnamefont {M.}~\bibnamefont
  {Rigol}}, \bibinfo {author} {\bibfnamefont {V.}~\bibnamefont {Dunjko}},
  \bibinfo {author} {\bibfnamefont {V.}~\bibnamefont {Yurovsky}}, \ and\
  \bibinfo {author} {\bibfnamefont {M.}~\bibnamefont {Olshanii}},\ }\href
  {\doibase 10.1103/PhysRevLett.98.050405} {\bibfield  {journal} {\bibinfo
  {journal} {Phys. Rev. Lett.}\ }\textbf {\bibinfo {volume} {98}},\ \bibinfo
  {pages} {050405} (\bibinfo {year} {2007})}\BibitemShut {NoStop}%
\bibitem [{\citenamefont {Rigol}\ \emph {et~al.}(2006)\citenamefont {Rigol},
  \citenamefont {Muramatsu},\ and\ \citenamefont {Olshanii}}]{Rigol06}%
  \BibitemOpen
  \bibfield  {author} {\bibinfo {author} {\bibfnamefont {M.}~\bibnamefont
  {Rigol}}, \bibinfo {author} {\bibfnamefont {A.}~\bibnamefont {Muramatsu}}, \
  and\ \bibinfo {author} {\bibfnamefont {M.}~\bibnamefont {Olshanii}},\ }\href
  {\doibase 10.1103/PhysRevA.74.053616} {\bibfield  {journal} {\bibinfo
  {journal} {Phys. Rev. A}\ }\textbf {\bibinfo {volume} {74}},\ \bibinfo
  {pages} {053616} (\bibinfo {year} {2006})}\BibitemShut {NoStop}%
\bibitem [{\citenamefont {Cazalilla}(2006)}]{Cazalilla06}%
  \BibitemOpen
  \bibfield  {author} {\bibinfo {author} {\bibfnamefont {M.~A.}\ \bibnamefont
  {Cazalilla}},\ }\href {\doibase 10.1103/PhysRevLett.97.156403} {\bibfield
  {journal} {\bibinfo  {journal} {Phys. Rev. Lett.}\ }\textbf {\bibinfo
  {volume} {97}},\ \bibinfo {pages} {156403} (\bibinfo {year}
  {2006})}\BibitemShut {NoStop}%
\bibitem [{\citenamefont {Iucci}\ and\ \citenamefont
  {Cazalilla}(2009)}]{Iucci09}%
  \BibitemOpen
  \bibfield  {author} {\bibinfo {author} {\bibfnamefont {A.}~\bibnamefont
  {Iucci}}\ and\ \bibinfo {author} {\bibfnamefont {M.~A.}\ \bibnamefont
  {Cazalilla}},\ }\href {\doibase 10.1103/PhysRevA.80.063619} {\bibfield
  {journal} {\bibinfo  {journal} {Phys. Rev. A}\ }\textbf {\bibinfo {volume}
  {80}},\ \bibinfo {pages} {063619} (\bibinfo {year} {2009})}\BibitemShut
  {NoStop}%
\bibitem [{\citenamefont {Kollar}\ and\ \citenamefont
  {Eckstein}(2008)}]{Kollar08}%
  \BibitemOpen
  \bibfield  {author} {\bibinfo {author} {\bibfnamefont {M.}~\bibnamefont
  {Kollar}}\ and\ \bibinfo {author} {\bibfnamefont {M.}~\bibnamefont
  {Eckstein}},\ }\href {\doibase 10.1103/PhysRevA.78.013626} {\bibfield
  {journal} {\bibinfo  {journal} {Phys. Rev. A}\ }\textbf {\bibinfo {volume}
  {78}},\ \bibinfo {pages} {013626} (\bibinfo {year} {2008})}\BibitemShut
  {NoStop}%
\bibitem [{\citenamefont {Eckstein}\ and\ \citenamefont
  {Kollar}(2008)}]{Eckstein08}%
  \BibitemOpen
  \bibfield  {author} {\bibinfo {author} {\bibfnamefont {M.}~\bibnamefont
  {Eckstein}}\ and\ \bibinfo {author} {\bibfnamefont {M.}~\bibnamefont
  {Kollar}},\ }\href {\doibase 10.1103/PhysRevLett.100.120404} {\bibfield
  {journal} {\bibinfo  {journal} {Phys. Rev. Lett.}\ }\textbf {\bibinfo
  {volume} {100}},\ \bibinfo {pages} {120404} (\bibinfo {year}
  {2008})}\BibitemShut {NoStop}%
\bibitem [{\citenamefont {Fioretto}\ and\ \citenamefont
  {Mussardo}(2010)}]{Fioretto_2010}%
  \BibitemOpen
  \bibfield  {author} {\bibinfo {author} {\bibfnamefont {D.}~\bibnamefont
  {Fioretto}}\ and\ \bibinfo {author} {\bibfnamefont {G.}~\bibnamefont
  {Mussardo}},\ }\href {\doibase 10.1088/1367-2630/12/5/055015} {\bibfield
  {journal} {\bibinfo  {journal} {New Journal of Physics}\ }\textbf {\bibinfo
  {volume} {12}},\ \bibinfo {pages} {055015} (\bibinfo {year}
  {2010})}\BibitemShut {NoStop}%
\bibitem [{\citenamefont {Roux}(2009)}]{Roux09}%
  \BibitemOpen
  \bibfield  {author} {\bibinfo {author} {\bibfnamefont {G.}~\bibnamefont
  {Roux}},\ }\href {\doibase 10.1103/PhysRevA.79.021608} {\bibfield  {journal}
  {\bibinfo  {journal} {Phys. Rev. A}\ }\textbf {\bibinfo {volume} {79}},\
  \bibinfo {pages} {021608} (\bibinfo {year} {2009})}\BibitemShut {NoStop}%
\bibitem [{\citenamefont {Kollath}\ \emph {et~al.}(2007)\citenamefont
  {Kollath}, \citenamefont {L\"auchli},\ and\ \citenamefont
  {Altman}}]{Kollath07}%
  \BibitemOpen
  \bibfield  {author} {\bibinfo {author} {\bibfnamefont {C.}~\bibnamefont
  {Kollath}}, \bibinfo {author} {\bibfnamefont {A.~M.}\ \bibnamefont
  {L\"auchli}}, \ and\ \bibinfo {author} {\bibfnamefont {E.}~\bibnamefont
  {Altman}},\ }\href {\doibase 10.1103/PhysRevLett.98.180601} {\bibfield
  {journal} {\bibinfo  {journal} {Phys. Rev. Lett.}\ }\textbf {\bibinfo
  {volume} {98}},\ \bibinfo {pages} {180601} (\bibinfo {year}
  {2007})}\BibitemShut {NoStop}%
\bibitem [{\citenamefont {Moeckel}\ and\ \citenamefont
  {Kehrein}(2008)}]{Moeckel08}%
  \BibitemOpen
  \bibfield  {author} {\bibinfo {author} {\bibfnamefont {M.}~\bibnamefont
  {Moeckel}}\ and\ \bibinfo {author} {\bibfnamefont {S.}~\bibnamefont
  {Kehrein}},\ }\href {\doibase 10.1103/PhysRevLett.100.175702} {\bibfield
  {journal} {\bibinfo  {journal} {Phys. Rev. Lett.}\ }\textbf {\bibinfo
  {volume} {100}},\ \bibinfo {pages} {175702} (\bibinfo {year}
  {2008})}\BibitemShut {NoStop}%
\bibitem [{\citenamefont {Moeckel}\ and\ \citenamefont
  {Kehrein}(2009)}]{Moeckel09}%
  \BibitemOpen
  \bibfield  {author} {\bibinfo {author} {\bibfnamefont {M.}~\bibnamefont
  {Moeckel}}\ and\ \bibinfo {author} {\bibfnamefont {S.}~\bibnamefont
  {Kehrein}},\ }\href {\doibase https://doi.org/10.1016/j.aop.2009.03.009}
  {\bibfield  {journal} {\bibinfo  {journal} {Annals of Physics}\ }\textbf
  {\bibinfo {volume} {324}},\ \bibinfo {pages} {2146 } (\bibinfo {year}
  {2009})}\BibitemShut {NoStop}%
\bibitem [{\citenamefont {Rigol}(2009)}]{Rigol09}%
  \BibitemOpen
  \bibfield  {author} {\bibinfo {author} {\bibfnamefont {M.}~\bibnamefont
  {Rigol}},\ }\href {\doibase 10.1103/PhysRevLett.103.100403} {\bibfield
  {journal} {\bibinfo  {journal} {Phys. Rev. Lett.}\ }\textbf {\bibinfo
  {volume} {103}},\ \bibinfo {pages} {100403} (\bibinfo {year}
  {2009})}\BibitemShut {NoStop}%
\bibitem [{\citenamefont {Rohwer}\ \emph {et~al.}(2011)\citenamefont {Rohwer},
  \citenamefont {Hellmann}, \citenamefont {Wiesenmayer}, \citenamefont {Sohrt},
  \citenamefont {Stange}, \citenamefont {Slomski}, \citenamefont {Carr},
  \citenamefont {Liu}, \citenamefont {Avila}, \citenamefont {Kall{\"{a}}ne},
  \citenamefont {Mathias}, \citenamefont {Kipp}, \citenamefont {Rossnagel},\
  and\ \citenamefont {Bauer}}]{Rohwer2011}%
  \BibitemOpen
  \bibfield  {author} {\bibinfo {author} {\bibfnamefont {T.}~\bibnamefont
  {Rohwer}}, \bibinfo {author} {\bibfnamefont {S.}~\bibnamefont {Hellmann}},
  \bibinfo {author} {\bibfnamefont {M.}~\bibnamefont {Wiesenmayer}}, \bibinfo
  {author} {\bibfnamefont {C.}~\bibnamefont {Sohrt}}, \bibinfo {author}
  {\bibfnamefont {A.}~\bibnamefont {Stange}}, \bibinfo {author} {\bibfnamefont
  {B.}~\bibnamefont {Slomski}}, \bibinfo {author} {\bibfnamefont
  {A.}~\bibnamefont {Carr}}, \bibinfo {author} {\bibfnamefont {Y.}~\bibnamefont
  {Liu}}, \bibinfo {author} {\bibfnamefont {L.~M.}\ \bibnamefont {Avila}},
  \bibinfo {author} {\bibfnamefont {M.}~\bibnamefont {Kall{\"{a}}ne}}, \bibinfo
  {author} {\bibfnamefont {S.}~\bibnamefont {Mathias}}, \bibinfo {author}
  {\bibfnamefont {L.}~\bibnamefont {Kipp}}, \bibinfo {author} {\bibfnamefont
  {K.}~\bibnamefont {Rossnagel}}, \ and\ \bibinfo {author} {\bibfnamefont
  {M.}~\bibnamefont {Bauer}},\ }\href {\doibase 10.1038/nature09829} {\bibfield
   {journal} {\bibinfo  {journal} {Nature}\ }\textbf {\bibinfo {volume}
  {471}},\ \bibinfo {pages} {490} (\bibinfo {year} {2011})}\BibitemShut
  {NoStop}%
\bibitem [{\citenamefont {Petersen}\ \emph {et~al.}(2011)\citenamefont
  {Petersen}, \citenamefont {Kaiser}, \citenamefont {Dean}, \citenamefont
  {Simoncig}, \citenamefont {Liu}, \citenamefont {Cavalieri}, \citenamefont
  {Cacho}, \citenamefont {Turcu}, \citenamefont {Springate}, \citenamefont
  {Frassetto}, \citenamefont {Poletto}, \citenamefont {Dhesi}, \citenamefont
  {Berger},\ and\ \citenamefont {Cavalleri}}]{Petersen11}%
  \BibitemOpen
  \bibfield  {author} {\bibinfo {author} {\bibfnamefont {J.~C.}\ \bibnamefont
  {Petersen}}, \bibinfo {author} {\bibfnamefont {S.}~\bibnamefont {Kaiser}},
  \bibinfo {author} {\bibfnamefont {N.}~\bibnamefont {Dean}}, \bibinfo {author}
  {\bibfnamefont {A.}~\bibnamefont {Simoncig}}, \bibinfo {author}
  {\bibfnamefont {H.~Y.}\ \bibnamefont {Liu}}, \bibinfo {author} {\bibfnamefont
  {A.~L.}\ \bibnamefont {Cavalieri}}, \bibinfo {author} {\bibfnamefont
  {C.}~\bibnamefont {Cacho}}, \bibinfo {author} {\bibfnamefont {I.~C.~E.}\
  \bibnamefont {Turcu}}, \bibinfo {author} {\bibfnamefont {E.}~\bibnamefont
  {Springate}}, \bibinfo {author} {\bibfnamefont {F.}~\bibnamefont
  {Frassetto}}, \bibinfo {author} {\bibfnamefont {L.}~\bibnamefont {Poletto}},
  \bibinfo {author} {\bibfnamefont {S.~S.}\ \bibnamefont {Dhesi}}, \bibinfo
  {author} {\bibfnamefont {H.}~\bibnamefont {Berger}}, \ and\ \bibinfo {author}
  {\bibfnamefont {A.}~\bibnamefont {Cavalleri}},\ }\href {\doibase
  10.1103/PhysRevLett.107.177402} {\bibfield  {journal} {\bibinfo  {journal}
  {Phys. Rev. Lett.}\ }\textbf {\bibinfo {volume} {107}},\ \bibinfo {pages}
  {177402} (\bibinfo {year} {2011})}\BibitemShut {NoStop}%
\bibitem [{\citenamefont {Perfetti}\ \emph {et~al.}(2006)\citenamefont
  {Perfetti}, \citenamefont {Loukakos}, \citenamefont {Lisowski}, \citenamefont
  {Bovensiepen}, \citenamefont {Berger}, \citenamefont {Biermann},
  \citenamefont {Cornaglia}, \citenamefont {Georges},\ and\ \citenamefont
  {Wolf}}]{Perfetti06}%
  \BibitemOpen
  \bibfield  {author} {\bibinfo {author} {\bibfnamefont {L.}~\bibnamefont
  {Perfetti}}, \bibinfo {author} {\bibfnamefont {P.~A.}\ \bibnamefont
  {Loukakos}}, \bibinfo {author} {\bibfnamefont {M.}~\bibnamefont {Lisowski}},
  \bibinfo {author} {\bibfnamefont {U.}~\bibnamefont {Bovensiepen}}, \bibinfo
  {author} {\bibfnamefont {H.}~\bibnamefont {Berger}}, \bibinfo {author}
  {\bibfnamefont {S.}~\bibnamefont {Biermann}}, \bibinfo {author}
  {\bibfnamefont {P.~S.}\ \bibnamefont {Cornaglia}}, \bibinfo {author}
  {\bibfnamefont {A.}~\bibnamefont {Georges}}, \ and\ \bibinfo {author}
  {\bibfnamefont {M.}~\bibnamefont {Wolf}},\ }\href {\doibase
  10.1103/PhysRevLett.97.067402} {\bibfield  {journal} {\bibinfo  {journal}
  {Phys. Rev. Lett.}\ }\textbf {\bibinfo {volume} {97}},\ \bibinfo {pages}
  {067402} (\bibinfo {year} {2006})}\BibitemShut {NoStop}%
\bibitem [{\citenamefont {Perfetti}\ \emph {et~al.}(2008)\citenamefont
  {Perfetti}, \citenamefont {Loukakos}, \citenamefont {Lisowski}, \citenamefont
  {Bovensiepen}, \citenamefont {Wolf}, \citenamefont {Berger}, \citenamefont
  {Biermann},\ and\ \citenamefont {Georges}}]{Perfetti_2008}%
  \BibitemOpen
  \bibfield  {author} {\bibinfo {author} {\bibfnamefont {L.}~\bibnamefont
  {Perfetti}}, \bibinfo {author} {\bibfnamefont {P.~A.}\ \bibnamefont
  {Loukakos}}, \bibinfo {author} {\bibfnamefont {M.}~\bibnamefont {Lisowski}},
  \bibinfo {author} {\bibfnamefont {U.}~\bibnamefont {Bovensiepen}}, \bibinfo
  {author} {\bibfnamefont {M.}~\bibnamefont {Wolf}}, \bibinfo {author}
  {\bibfnamefont {H.}~\bibnamefont {Berger}}, \bibinfo {author} {\bibfnamefont
  {S.}~\bibnamefont {Biermann}}, \ and\ \bibinfo {author} {\bibfnamefont
  {A.}~\bibnamefont {Georges}},\ }\href {\doibase
  10.1088/1367-2630/10/5/053019} {\bibfield  {journal} {\bibinfo  {journal}
  {New Journal of Physics}\ }\textbf {\bibinfo {volume} {10}},\ \bibinfo
  {pages} {053019} (\bibinfo {year} {2008})}\BibitemShut {NoStop}%
\bibitem [{\citenamefont {Avigo}\ \emph {et~al.}(2016)\citenamefont {Avigo},
  \citenamefont {Thirupathaiah}, \citenamefont {Ligges}, \citenamefont {Wolf},
  \citenamefont {Fink},\ and\ \citenamefont {Bovensiepen}}]{Avigo_2016}%
  \BibitemOpen
  \bibfield  {author} {\bibinfo {author} {\bibfnamefont {I.}~\bibnamefont
  {Avigo}}, \bibinfo {author} {\bibfnamefont {S.}~\bibnamefont
  {Thirupathaiah}}, \bibinfo {author} {\bibfnamefont {M.}~\bibnamefont
  {Ligges}}, \bibinfo {author} {\bibfnamefont {T.}~\bibnamefont {Wolf}},
  \bibinfo {author} {\bibfnamefont {J.}~\bibnamefont {Fink}}, \ and\ \bibinfo
  {author} {\bibfnamefont {U.}~\bibnamefont {Bovensiepen}},\ }\href {\doibase
  10.1088/1367-2630/18/9/093028} {\bibfield  {journal} {\bibinfo  {journal}
  {New Journal of Physics}\ }\textbf {\bibinfo {volume} {18}},\ \bibinfo
  {pages} {093028} (\bibinfo {year} {2016})}\BibitemShut {NoStop}%
\bibitem [{\citenamefont {Wu}\ \emph {et~al.}(2021)\citenamefont {Wu},
  \citenamefont {Wang}, \citenamefont {Yang}, \citenamefont {Duan},
  \citenamefont {Huang}, \citenamefont {Tang}, \citenamefont {Guo},
  \citenamefont {Luo},\ and\ \citenamefont {Zhang}}]{Wu_2021}%
  \BibitemOpen
  \bibfield  {author} {\bibinfo {author} {\bibfnamefont {T.}~\bibnamefont
  {Wu}}, \bibinfo {author} {\bibfnamefont {H.}~\bibnamefont {Wang}}, \bibinfo
  {author} {\bibfnamefont {Y.}~\bibnamefont {Yang}}, \bibinfo {author}
  {\bibfnamefont {S.}~\bibnamefont {Duan}}, \bibinfo {author} {\bibfnamefont
  {C.}~\bibnamefont {Huang}}, \bibinfo {author} {\bibfnamefont
  {T.}~\bibnamefont {Tang}}, \bibinfo {author} {\bibfnamefont {Y.}~\bibnamefont
  {Guo}}, \bibinfo {author} {\bibfnamefont {W.}~\bibnamefont {Luo}}, \ and\
  \bibinfo {author} {\bibfnamefont {W.}~\bibnamefont {Zhang}},\ }\href
  {\doibase 10.1088/1674-1056/ac373c} {\bibfield  {journal} {\bibinfo
  {journal} {Chinese Physics B}\ }\textbf {\bibinfo {volume} {31}},\ \bibinfo
  {pages} {027902} (\bibinfo {year} {2021})}\BibitemShut {NoStop}%
\bibitem [{\citenamefont {Schmitt}\ \emph {et~al.}(2011)\citenamefont
  {Schmitt}, \citenamefont {Kirchmann}, \citenamefont {Bovensiepen},
  \citenamefont {Moore}, \citenamefont {Chu}, \citenamefont {Lu}, \citenamefont
  {Rettig}, \citenamefont {Wolf}, \citenamefont {Fisher},\ and\ \citenamefont
  {Shen}}]{Schmitt_2011}%
  \BibitemOpen
  \bibfield  {author} {\bibinfo {author} {\bibfnamefont {F.}~\bibnamefont
  {Schmitt}}, \bibinfo {author} {\bibfnamefont {P.~S.}\ \bibnamefont
  {Kirchmann}}, \bibinfo {author} {\bibfnamefont {U.}~\bibnamefont
  {Bovensiepen}}, \bibinfo {author} {\bibfnamefont {R.~G.}\ \bibnamefont
  {Moore}}, \bibinfo {author} {\bibfnamefont {J.-H.}\ \bibnamefont {Chu}},
  \bibinfo {author} {\bibfnamefont {D.~H.}\ \bibnamefont {Lu}}, \bibinfo
  {author} {\bibfnamefont {L.}~\bibnamefont {Rettig}}, \bibinfo {author}
  {\bibfnamefont {M.}~\bibnamefont {Wolf}}, \bibinfo {author} {\bibfnamefont
  {I.~R.}\ \bibnamefont {Fisher}}, \ and\ \bibinfo {author} {\bibfnamefont
  {Z.-X.}\ \bibnamefont {Shen}},\ }\href {\doibase
  10.1088/1367-2630/13/6/063022} {\bibfield  {journal} {\bibinfo  {journal}
  {New Journal of Physics}\ }\textbf {\bibinfo {volume} {13}},\ \bibinfo
  {pages} {063022} (\bibinfo {year} {2011})}\BibitemShut {NoStop}%
\bibitem [{\citenamefont {Aoki}\ \emph {et~al.}(2014)\citenamefont {Aoki},
  \citenamefont {Tsuji}, \citenamefont {Eckstein}, \citenamefont {Kollar},
  \citenamefont {Oka},\ and\ \citenamefont {Werner}}]{Aoki14}%
  \BibitemOpen
  \bibfield  {author} {\bibinfo {author} {\bibfnamefont {H.}~\bibnamefont
  {Aoki}}, \bibinfo {author} {\bibfnamefont {N.}~\bibnamefont {Tsuji}},
  \bibinfo {author} {\bibfnamefont {M.}~\bibnamefont {Eckstein}}, \bibinfo
  {author} {\bibfnamefont {M.}~\bibnamefont {Kollar}}, \bibinfo {author}
  {\bibfnamefont {T.}~\bibnamefont {Oka}}, \ and\ \bibinfo {author}
  {\bibfnamefont {P.}~\bibnamefont {Werner}},\ }\href {\doibase
  10.1103/RevModPhys.86.779} {\bibfield  {journal} {\bibinfo  {journal} {Rev.
  Mod. Phys.}\ }\textbf {\bibinfo {volume} {86}},\ \bibinfo {pages} {779}
  (\bibinfo {year} {2014})}\BibitemShut {NoStop}%
\bibitem [{\citenamefont {Nosarzewski}\ \emph {et~al.}(2017)\citenamefont
  {Nosarzewski}, \citenamefont {Moritz}, \citenamefont {Freericks},
  \citenamefont {Kemper},\ and\ \citenamefont {Devereaux}}]{Nosarzewski17}%
  \BibitemOpen
  \bibfield  {author} {\bibinfo {author} {\bibfnamefont {B.}~\bibnamefont
  {Nosarzewski}}, \bibinfo {author} {\bibfnamefont {B.}~\bibnamefont {Moritz}},
  \bibinfo {author} {\bibfnamefont {J.~K.}\ \bibnamefont {Freericks}}, \bibinfo
  {author} {\bibfnamefont {A.~F.}\ \bibnamefont {Kemper}}, \ and\ \bibinfo
  {author} {\bibfnamefont {T.~P.}\ \bibnamefont {Devereaux}},\ }\href {\doibase
  10.1103/PhysRevB.96.184518} {\bibfield  {journal} {\bibinfo  {journal} {Phys.
  Rev. B}\ }\textbf {\bibinfo {volume} {96}},\ \bibinfo {pages} {184518}
  (\bibinfo {year} {2017})}\BibitemShut {NoStop}%
\bibitem [{\citenamefont {Kemper}\ \emph {et~al.}(2015)\citenamefont {Kemper},
  \citenamefont {Sentef}, \citenamefont {Moritz}, \citenamefont {Freericks},\
  and\ \citenamefont {Devereaux}}]{Kemper15}%
  \BibitemOpen
  \bibfield  {author} {\bibinfo {author} {\bibfnamefont {A.~F.}\ \bibnamefont
  {Kemper}}, \bibinfo {author} {\bibfnamefont {M.~A.}\ \bibnamefont {Sentef}},
  \bibinfo {author} {\bibfnamefont {B.}~\bibnamefont {Moritz}}, \bibinfo
  {author} {\bibfnamefont {J.~K.}\ \bibnamefont {Freericks}}, \ and\ \bibinfo
  {author} {\bibfnamefont {T.~P.}\ \bibnamefont {Devereaux}},\ }\href {\doibase
  10.1103/PhysRevB.92.224517} {\bibfield  {journal} {\bibinfo  {journal} {Phys.
  Rev. B}\ }\textbf {\bibinfo {volume} {92}},\ \bibinfo {pages} {224517}
  (\bibinfo {year} {2015})}\BibitemShut {NoStop}%
\bibitem [{\citenamefont {Kemper}\ \emph {et~al.}(2017)\citenamefont {Kemper},
  \citenamefont {Sentef}, \citenamefont {Moritz}, \citenamefont {Devereaux},\
  and\ \citenamefont {Freericks}}]{Kemper17}%
  \BibitemOpen
  \bibfield  {author} {\bibinfo {author} {\bibfnamefont {A.~F.}\ \bibnamefont
  {Kemper}}, \bibinfo {author} {\bibfnamefont {M.~A.}\ \bibnamefont {Sentef}},
  \bibinfo {author} {\bibfnamefont {B.}~\bibnamefont {Moritz}}, \bibinfo
  {author} {\bibfnamefont {T.~P.}\ \bibnamefont {Devereaux}}, \ and\ \bibinfo
  {author} {\bibfnamefont {J.~K.}\ \bibnamefont {Freericks}},\ }\href {\doibase
  10.1002/andp.201600235} {\bibfield  {journal} {\bibinfo  {journal} {Annalen
  der Physik}\ }\textbf {\bibinfo {volume} {529}},\ \bibinfo {pages} {n/a}
  (\bibinfo {year} {2017})}\BibitemShut {NoStop}%
\bibitem [{\citenamefont {Shao}\ \emph {et~al.}(2022)\citenamefont {Shao},
  \citenamefont {Lu}, \citenamefont {Zhang}, \citenamefont {Yu}, \citenamefont
  {Tohyama},\ and\ \citenamefont {Lu}}]{ShaoCan2022}%
  \BibitemOpen
  \bibfield  {author} {\bibinfo {author} {\bibfnamefont {C.}~\bibnamefont
  {Shao}}, \bibinfo {author} {\bibfnamefont {H.}~\bibnamefont {Lu}}, \bibinfo
  {author} {\bibfnamefont {X.}~\bibnamefont {Zhang}}, \bibinfo {author}
  {\bibfnamefont {C.}~\bibnamefont {Yu}}, \bibinfo {author} {\bibfnamefont
  {T.}~\bibnamefont {Tohyama}}, \ and\ \bibinfo {author} {\bibfnamefont
  {R.}~\bibnamefont {Lu}},\ }\href {\doibase 10.1103/PhysRevLett.128.047401}
  {\bibfield  {journal} {\bibinfo  {journal} {Phys. Rev. Lett.}\ }\textbf
  {\bibinfo {volume} {128}},\ \bibinfo {pages} {047401} (\bibinfo {year}
  {2022})}\BibitemShut {NoStop}%
\bibitem [{\citenamefont {Shao}\ \emph {et~al.}(2019)\citenamefont {Shao},
  \citenamefont {Lu}, \citenamefont {Luo},\ and\ \citenamefont
  {Mondaini}}]{ShaoCan2019}%
  \BibitemOpen
  \bibfield  {author} {\bibinfo {author} {\bibfnamefont {C.}~\bibnamefont
  {Shao}}, \bibinfo {author} {\bibfnamefont {H.}~\bibnamefont {Lu}}, \bibinfo
  {author} {\bibfnamefont {H.-G.}\ \bibnamefont {Luo}}, \ and\ \bibinfo
  {author} {\bibfnamefont {R.}~\bibnamefont {Mondaini}},\ }\href {\doibase
  10.1103/PhysRevB.100.041114} {\bibfield  {journal} {\bibinfo  {journal}
  {Phys. Rev. B}\ }\textbf {\bibinfo {volume} {100}},\ \bibinfo {pages}
  {041114} (\bibinfo {year} {2019})}\BibitemShut {NoStop}%
\bibitem [{\citenamefont {Shao}\ \emph {et~al.}(2016)\citenamefont {Shao},
  \citenamefont {Tohyama}, \citenamefont {Luo},\ and\ \citenamefont
  {Lu}}]{ShaoCan2016}%
  \BibitemOpen
  \bibfield  {author} {\bibinfo {author} {\bibfnamefont {C.}~\bibnamefont
  {Shao}}, \bibinfo {author} {\bibfnamefont {T.}~\bibnamefont {Tohyama}},
  \bibinfo {author} {\bibfnamefont {H.-G.}\ \bibnamefont {Luo}}, \ and\
  \bibinfo {author} {\bibfnamefont {H.}~\bibnamefont {Lu}},\ }\href {\doibase
  10.1103/PhysRevB.93.195144} {\bibfield  {journal} {\bibinfo  {journal} {Phys.
  Rev. B}\ }\textbf {\bibinfo {volume} {93}},\ \bibinfo {pages} {195144}
  (\bibinfo {year} {2016})}\BibitemShut {NoStop}%
\bibitem [{\citenamefont {Kanamori}\ \emph {et~al.}(2009)\citenamefont
  {Kanamori}, \citenamefont {Matsueda},\ and\ \citenamefont
  {Ishihara}}]{Kanamori09}%
  \BibitemOpen
  \bibfield  {author} {\bibinfo {author} {\bibfnamefont {Y.}~\bibnamefont
  {Kanamori}}, \bibinfo {author} {\bibfnamefont {H.}~\bibnamefont {Matsueda}},
  \ and\ \bibinfo {author} {\bibfnamefont {S.}~\bibnamefont {Ishihara}},\
  }\href {\doibase 10.1103/PhysRevLett.103.267401} {\bibfield  {journal}
  {\bibinfo  {journal} {Phys. Rev. Lett.}\ }\textbf {\bibinfo {volume} {103}},\
  \bibinfo {pages} {267401} (\bibinfo {year} {2009})}\BibitemShut {NoStop}%
\bibitem [{\citenamefont {Tsutsui}\ \emph {et~al.}(1996)\citenamefont
  {Tsutsui}, \citenamefont {Ohta}, \citenamefont {Eder}, \citenamefont
  {Maekawa}, \citenamefont {Dagotto},\ and\ \citenamefont {Riera}}]{Tsutsui96}%
  \BibitemOpen
  \bibfield  {author} {\bibinfo {author} {\bibfnamefont {K.}~\bibnamefont
  {Tsutsui}}, \bibinfo {author} {\bibfnamefont {Y.}~\bibnamefont {Ohta}},
  \bibinfo {author} {\bibfnamefont {R.}~\bibnamefont {Eder}}, \bibinfo {author}
  {\bibfnamefont {S.}~\bibnamefont {Maekawa}}, \bibinfo {author} {\bibfnamefont
  {E.}~\bibnamefont {Dagotto}}, \ and\ \bibinfo {author} {\bibfnamefont
  {J.}~\bibnamefont {Riera}},\ }\href {\doibase 10.1103/PhysRevLett.76.279}
  {\bibfield  {journal} {\bibinfo  {journal} {Phys. Rev. Lett.}\ }\textbf
  {\bibinfo {volume} {76}},\ \bibinfo {pages} {279} (\bibinfo {year}
  {1996})}\BibitemShut {NoStop}%
\bibitem [{\citenamefont {Tohyama}(2004)}]{Tohyama04}%
  \BibitemOpen
  \bibfield  {author} {\bibinfo {author} {\bibfnamefont {T.}~\bibnamefont
  {Tohyama}},\ }\href {\doibase 10.1103/PhysRevB.70.174517} {\bibfield
  {journal} {\bibinfo  {journal} {Phys. Rev. B}\ }\textbf {\bibinfo {volume}
  {70}},\ \bibinfo {pages} {174517} (\bibinfo {year} {2004})}\BibitemShut
  {NoStop}%
\bibitem [{\citenamefont {Shao}\ \emph {et~al.}(2020)\citenamefont {Shao},
  \citenamefont {Tohyama}, \citenamefont {Luo},\ and\ \citenamefont
  {Lu}}]{Shao20}%
  \BibitemOpen
  \bibfield  {author} {\bibinfo {author} {\bibfnamefont {C.}~\bibnamefont
  {Shao}}, \bibinfo {author} {\bibfnamefont {T.}~\bibnamefont {Tohyama}},
  \bibinfo {author} {\bibfnamefont {H.-G.}\ \bibnamefont {Luo}}, \ and\
  \bibinfo {author} {\bibfnamefont {H.}~\bibnamefont {Lu}},\ }\href {\doibase
  10.1103/PhysRevB.101.045128} {\bibfield  {journal} {\bibinfo  {journal}
  {Phys. Rev. B}\ }\textbf {\bibinfo {volume} {101}},\ \bibinfo {pages}
  {045128} (\bibinfo {year} {2020})}\BibitemShut {NoStop}%
\bibitem [{\citenamefont {Bohrdt}\ \emph {et~al.}(2018)\citenamefont {Bohrdt},
  \citenamefont {Greif}, \citenamefont {Demler}, \citenamefont {Knap},\ and\
  \citenamefont {Grusdt}}]{Bohrdt18}%
  \BibitemOpen
  \bibfield  {author} {\bibinfo {author} {\bibfnamefont {A.}~\bibnamefont
  {Bohrdt}}, \bibinfo {author} {\bibfnamefont {D.}~\bibnamefont {Greif}},
  \bibinfo {author} {\bibfnamefont {E.}~\bibnamefont {Demler}}, \bibinfo
  {author} {\bibfnamefont {M.}~\bibnamefont {Knap}}, \ and\ \bibinfo {author}
  {\bibfnamefont {F.}~\bibnamefont {Grusdt}},\ }\href {\doibase
  10.1103/PhysRevB.97.125117} {\bibfield  {journal} {\bibinfo  {journal} {Phys.
  Rev. B}\ }\textbf {\bibinfo {volume} {97}},\ \bibinfo {pages} {125117}
  (\bibinfo {year} {2018})}\BibitemShut {NoStop}%
\bibitem [{\citenamefont {Emery}()}]{Emery}%
  \BibitemOpen
  \bibfield  {author} {\bibinfo {author} {\bibfnamefont {V.~J.}\ \bibnamefont
  {Emery}},\ }\href@noop {} {\bibinfo  {journal} {in {\it Highly Conducting
  One-Dimensional Solids}, edited by J. Devreese, R. Evrard, and V. van Doren,
  (Plenum, New York, 1979), p. 247}\ }\BibitemShut {NoStop}%
\bibitem [{\citenamefont {Tsuchiizu}\ and\ \citenamefont
  {Furusaki}(2002)}]{Tsuchiizu02}%
  \BibitemOpen
\bibfield  {journal} {  }\bibfield  {author} {\bibinfo {author} {\bibfnamefont
  {M.}~\bibnamefont {Tsuchiizu}}\ and\ \bibinfo {author} {\bibfnamefont
  {A.}~\bibnamefont {Furusaki}},\ }\href {\doibase
  10.1103/PhysRevLett.88.056402} {\bibfield  {journal} {\bibinfo  {journal}
  {Phys. Rev. Lett.}\ }\textbf {\bibinfo {volume} {88}},\ \bibinfo {pages}
  {056402} (\bibinfo {year} {2002})}\BibitemShut {NoStop}%
\bibitem [{\citenamefont {Ejima}\ and\ \citenamefont
  {Nishimoto}(2007)}]{Ejima07}%
  \BibitemOpen
  \bibfield  {author} {\bibinfo {author} {\bibfnamefont {S.}~\bibnamefont
  {Ejima}}\ and\ \bibinfo {author} {\bibfnamefont {S.}~\bibnamefont
  {Nishimoto}},\ }\href {\doibase 10.1103/PhysRevLett.99.216403} {\bibfield
  {journal} {\bibinfo  {journal} {Phys. Rev. Lett.}\ }\textbf {\bibinfo
  {volume} {99}},\ \bibinfo {pages} {216403} (\bibinfo {year}
  {2007})}\BibitemShut {NoStop}%
\bibitem [{\citenamefont {Poilblanc}(1991)}]{Poilblanc91}%
  \BibitemOpen
  \bibfield  {author} {\bibinfo {author} {\bibfnamefont {D.}~\bibnamefont
  {Poilblanc}},\ }\href {\doibase 10.1103/PhysRevB.44.9562} {\bibfield
  {journal} {\bibinfo  {journal} {Phys. Rev. B}\ }\textbf {\bibinfo {volume}
  {44}},\ \bibinfo {pages} {9562} (\bibinfo {year} {1991})}\BibitemShut
  {NoStop}%
\bibitem [{\citenamefont {Prelov$\check{\text{s}}$ek}\ and\ \citenamefont
  {Bon$\check{\text{c}}$a}()}]{Prelovsek}%
  \BibitemOpen
  \bibfield  {author} {\bibinfo {author} {\bibfnamefont {P.}~\bibnamefont
  {Prelov$\check{\text{s}}$ek}}\ and\ \bibinfo {author} {\bibfnamefont
  {J.}~\bibnamefont {Bon$\check{\text{c}}$a}},\ }\href@noop {} {\bibinfo
  {journal} {in {\it Strongly Correlated Systems-Numerical Methods}, edited by
  A. Avella and F. Mancini, Springer Series in Solid-State Sciences, Vol. 176
  (Springer, Berlin, 2013), pp. 1-30}\ }\BibitemShut {NoStop}%
\bibitem [{\citenamefont {Kim}\ \emph {et~al.}(1996)\citenamefont {Kim},
  \citenamefont {Matsuura}, \citenamefont {Shen}, \citenamefont {Motoyama},
  \citenamefont {Eisaki}, \citenamefont {Uchida}, \citenamefont {Tohyama},\
  and\ \citenamefont {Maekawa}}]{Kim96}%
  \BibitemOpen
\bibfield  {journal} {  }\bibfield  {author} {\bibinfo {author} {\bibfnamefont
  {C.}~\bibnamefont {Kim}}, \bibinfo {author} {\bibfnamefont {A.~Y.}\
  \bibnamefont {Matsuura}}, \bibinfo {author} {\bibfnamefont {Z.-X.}\
  \bibnamefont {Shen}}, \bibinfo {author} {\bibfnamefont {N.}~\bibnamefont
  {Motoyama}}, \bibinfo {author} {\bibfnamefont {H.}~\bibnamefont {Eisaki}},
  \bibinfo {author} {\bibfnamefont {S.}~\bibnamefont {Uchida}}, \bibinfo
  {author} {\bibfnamefont {T.}~\bibnamefont {Tohyama}}, \ and\ \bibinfo
  {author} {\bibfnamefont {S.}~\bibnamefont {Maekawa}},\ }\href {\doibase
  10.1103/PhysRevLett.77.4054} {\bibfield  {journal} {\bibinfo  {journal}
  {Phys. Rev. Lett.}\ }\textbf {\bibinfo {volume} {77}},\ \bibinfo {pages}
  {4054} (\bibinfo {year} {1996})}\BibitemShut {NoStop}%
\bibitem [{\citenamefont {Aichhorn}\ \emph {et~al.}(2004)\citenamefont
  {Aichhorn}, \citenamefont {Evertz}, \citenamefont {von~der Linden},\ and\
  \citenamefont {Potthoff}}]{Aichhorn04}%
  \BibitemOpen
  \bibfield  {author} {\bibinfo {author} {\bibfnamefont {M.}~\bibnamefont
  {Aichhorn}}, \bibinfo {author} {\bibfnamefont {H.~G.}\ \bibnamefont
  {Evertz}}, \bibinfo {author} {\bibfnamefont {W.}~\bibnamefont {von~der
  Linden}}, \ and\ \bibinfo {author} {\bibfnamefont {M.}~\bibnamefont
  {Potthoff}},\ }\href {\doibase 10.1103/PhysRevB.70.235107} {\bibfield
  {journal} {\bibinfo  {journal} {Phys. Rev. B}\ }\textbf {\bibinfo {volume}
  {70}},\ \bibinfo {pages} {235107} (\bibinfo {year} {2004})}\BibitemShut
  {NoStop}%
\bibitem [{\citenamefont {Kim}\ \emph {et~al.}(2006)\citenamefont {Kim},
  \citenamefont {Koh}, \citenamefont {Rotenberg}, \citenamefont {Oh},
  \citenamefont {Eisaki}, \citenamefont {Motoyama}, \citenamefont {Uchida},
  \citenamefont {Tohyama}, \citenamefont {Maekawa}, \citenamefont {Shen},\ and\
  \citenamefont {Kim}}]{Kim2006}%
  \BibitemOpen
  \bibfield  {author} {\bibinfo {author} {\bibfnamefont {B.~J.}\ \bibnamefont
  {Kim}}, \bibinfo {author} {\bibfnamefont {H.}~\bibnamefont {Koh}}, \bibinfo
  {author} {\bibfnamefont {E.}~\bibnamefont {Rotenberg}}, \bibinfo {author}
  {\bibfnamefont {S.-J.}\ \bibnamefont {Oh}}, \bibinfo {author} {\bibfnamefont
  {H.}~\bibnamefont {Eisaki}}, \bibinfo {author} {\bibfnamefont
  {N.}~\bibnamefont {Motoyama}}, \bibinfo {author} {\bibfnamefont
  {S.}~\bibnamefont {Uchida}}, \bibinfo {author} {\bibfnamefont
  {T.}~\bibnamefont {Tohyama}}, \bibinfo {author} {\bibfnamefont
  {S.}~\bibnamefont {Maekawa}}, \bibinfo {author} {\bibfnamefont {Z.-X.}\
  \bibnamefont {Shen}}, \ and\ \bibinfo {author} {\bibfnamefont
  {C.}~\bibnamefont {Kim}},\ }\href {\doibase 10.1038/nphys316} {\bibfield
  {journal} {\bibinfo  {journal} {Nature Physics}\ }\textbf {\bibinfo {volume}
  {2}},\ \bibinfo {pages} {397} (\bibinfo {year} {2006})}\BibitemShut {NoStop}%
\bibitem [{\citenamefont {Jeckelmann}(2003)}]{Jeckelmann03}%
  \BibitemOpen
  \bibfield  {author} {\bibinfo {author} {\bibfnamefont {E.}~\bibnamefont
  {Jeckelmann}},\ }\href {\doibase 10.1103/PhysRevB.67.075106} {\bibfield
  {journal} {\bibinfo  {journal} {Phys. Rev. B}\ }\textbf {\bibinfo {volume}
  {67}},\ \bibinfo {pages} {075106} (\bibinfo {year} {2003})}\BibitemShut
  {NoStop}%
\end{thebibliography}

%

\end{document}